\documentclass[%
 reprint,
 amsmath,amssymb,
 aps,
prx,
nourl,noeprint,
]{revtex4-2}
\usepackage{geometry}                		% See geometry.pdf to learn the layout options. There are lots.	
\usepackage{dcolumn}
\usepackage{subfiles} % Best loaded last in the preamble
\usepackage{bm}
\usepackage{float}
\usepackage[utf8]{inputenc}
\usepackage{color}   %May be necessary if you want to color links
\usepackage{hyperref}
\hypersetup{
    colorlinks=true,
    linkcolor=blue,
    filecolor=magenta,      
    urlcolor=cyan,
}
\usepackage{graphicx} % Required for inserting images
\begin{document}

\preprint{APS/123-QED}

\title{Programming Shapes with Competing Layered Patterns}

\author{Wan Yee Yau}
\affiliation{Max Planck Institute of Molecular Cell Biology and Genetics, Dresden, Germany}
\affiliation{Center for Systems Biology Dresden, Dresden, Germany}
\affiliation{Cluster of Excellence, Physics of Life, TU Dresden, Dresden, Germany}

\author{Carl D. Modes}
 \email{modes@mpi-cbg.de}
\affiliation{Max Planck Institute of Molecular Cell Biology and Genetics, Dresden, Germany}
\affiliation{Center for Systems Biology Dresden, Dresden, Germany}
\affiliation{Cluster of Excellence, Physics of Life, TU Dresden, Dresden, Germany}

\date{\today}

\begin{abstract}
Studying shape changing thick surfaces induced by differential growth helps us understand morphogenesis in biology and offers opportunities for device design. 
While ideal 2D differential growth maps have been well studied for both isotropic and anisotropic growth, scenarios involving gradients in thickness growth are far less explored. 
In this paper, we focus on a bilayer system in which the two layers undergo independent but incompatible growth. We examine how the strength of the growth patterns and the aspect ratio of the bilayer influence the resulting shapes. 
%The resulting configurations exhibit significant residual strain, making the model more realistic for describing tissue growth in animals. 
%Another advantage of programming shapes in this setting is that complex geometries can be achieved through relatively simple deformation patterns on the bilayer.
%To understand the mechanism by which final configurations emerge in an incompatible bilayer, we decompose the problem into three steps. 
We first investigate the effect of global area difference in the bilayer. %to determine how stretching and bending energies depend on aspect ratio and growth mismatch. 
Next, we make one of the two layers active and program it with positive or negative, localized or uniform curvature. %to study how curvature sign and distribution influence the resulting shapes. 
We then present examples involving competition between two active surfaces with opposite curvature signs or different curvature distributions and understand how the final configurations follow from the principles identified earlier. 
Finally, we demonstrate the ability to program biologically inspired shapes, such as dental epithelium and the ventral furrow in Drosophila, using a bilayer with simple, distinct deformation gradients.
\end{abstract}

\maketitle

\section{Introduction}

Shape changes emerge naturally from biological systems and play an important role in the development and function of organisms.
Understanding these processes could provide practical uses in areas such as medical applications, tissue engineering, soft robotics, and 4D printing.

Taking inspiration from nature, we are here interested in systems with thick sheet geometry that deform into 3D shapes due to in-plane growth that may vary both across the sheet and through its thickness. Many systems in nature are believed to attain their shape in such a manner, such as leaves and petals in plants~\cite{Coen_2023,Wang2024-2}, as well as during epithelial morphogenesis in animals~\cite{Nelson2016,Thompson2019,Tomba2022}, which contributes to shaping important organs. 
We aim to understand the relationship between these in-plane growth patterns and their resulting 3D shapes, particularly in cases where the growth patterns are incompatible across the sheet thickness. 
%Incompatible patterns that vary smoothly through the thickness can also be thought of as an effective competition between the distinct patterns on each boundary surface or layer.
Such a competition between patterns may then give rise to complex 3D shapes from relatively simple growth profiles on the two layers. 
Furthermore, since biological tissues are often pre-stressed~\cite{Ciarletta2016}, studying incompatible growth in such systems may offer a more realistic and biologically relevant framework than models based solely on ideal 2D sheets with prescribed growth patterns or models based solely on apico-basal (i.e. through the tissue thickness) contractions.

The recent rise in appreciation of mechanics in tissue biology has lead to a better but still incomplete picture of shape acquisition in morphogenesis.
Indeed, a wide range of biological systems have been studied to understand the stresses and strains contributing to morphogenesis, including \textit{Drosophila} ventral furrow and wing disc formation \cite{Brodland_2010,Leptin_1990,Martin_2020,Sui_2018,Sweeton_1991,Fuhrmann2024}, chick intestine looping \cite{Huycke_2018}, brain gyrification \cite{Richman_1975,Akula_2023,Razavi_2015}, lung branching \cite{Herriges_2014,Kim_2015,Morrisey_2010}, and optic cup formation \cite{Ramos2025,Cardozo_2023,Casey_2021,Eiraku_2011}.
Morphogenesis in plants has also been investigated \cite{Coen_2023,Silveira_2025}, although the underlying mechanisms differ from those in animal systems owing to the presence in plant cells of a stiff cell wall. 

%In order to study these biological events and others like them, several modeling techniques have been developed and deployed.
%Vertex models, for example, represent epithelia as a set of vertices of a polygonal tiling of space, together with an energy function derived from the underlying biological mechanisms.
%2D apical vertex models can be used to investigate cell packing topology as well as the distribution of cell areas and the number of cell neighbours and their effects on tissue mechanics \cite{Alt_2016}.
%Extensions to 2D lateral vertex models and 3D vertex models have been developed to explore out-of-plane epithelial deformations \cite{Murisic_2015}.
%Alternatively, tissue can be treated as a continuous material within the framework of continuum mechanics and there has been considerable interest in analytic approaches here \cite{Munoz_2010,Khairy_2018}.
%Computationally, the finite element method has been widely applied to simulate continuum morphogenesis in various systems, such as ventral furrow formation in \textit{Drosophila} \cite{Conte_2007} and buckling in chick guts \cite{Gill_2024}.
%Spring lattice models have also been used, for example to model eversion in the \textit{Drosophila} wing disc. \cite{Fuhrmann2024}

Meanwhile, in non-biological physics and engineering, programming shape through material design has been extensively explored, particularly in nematic elastomer sheets and N-isopropylacrylamide (NIPA) gel and other hydrogel systems in 2D~\cite{Efrati2007}. 
In the case of the nematic solids, which undergo a uniaxial spontaneous strain upon actuation, a wide variety of 3D shapes can be achieved by controlling the director field in the flat configuration. 
Examples include introducing spiral fields leading to families of conical related shapes \cite{Mostajeran2016} or other field patterns giving rise to even larger families \cite{Duffy2021}. 
Several works have introduced topological defects into the 2D director field for increased stability of the director pattern and robustness of the shape control \cite{Modes2010, Modes2011-1, Modes2011-2, Kralj2021, Feng_2022, Duffy_2020}. 
In the case of the hydrogel systems, non-Euclidean metrics can be realized by introducing appropriate swelling gradients in disks \cite{Klein2007, Kim2012}. 
%For arbitrary target 3D shapes, conformal flattening methods can be used to numerically design the corresponding 2D in-plane swelling pattern \cite{Nojoomi_2021}.
%Experimentally, many such 3D shapes have been achieved using hydrogels with spatially controlled crosslinking profiles \cite{Klein2007, Kim2012}. 

There is less work focusing on differential growth where growth may also vary through the thickness.  
Several works have studied the bending of a surface of photoelastomers that has a gradient in response through its thickness.
In nematic sheets, it has been demonstrated that a gradient in the director field across the thickness can induce nontrivial hyperbolic curvature \cite{Aharoni2014}. An initially circular bilayer nematic elastomer disk with tangential and radial alignment of the director fields on each surface can spontaneously break circular symmetry and fold into a square \cite{Modes2012}. 
3D shapes can emerge from imposed strain mismatches achieved by uniaxially stretching the rubber layer in a rubber-plastic bilayer composite \cite{Ramachandran_2021}. Finally, several cases where one layer spontaneously undergoes uniform expansion while the other remains passive have also been studied \cite{Pezzulla_2016, Caruso_2018}. 

Van Rees et al. \cite{van_Rees_2017} introduced a solution to the inverse problem of bilayer systems under orthotropic growth, targeting shapes with maximum principal curvature below $3/2h$. Their approach is powerful for designing bilayers that deform into arbitrary curved surfaces, especially in thin sheets with no or very low residual stress after relaxation. However, the general forward problem, particularly for bilayers in the thick limit, remains unsolved. Here, we focus on the forward problem in thicker bilayer sheets, where a significant amount of residual spontaneous strain remains after relaxation, since residual stress is known in many cases to play an important role in biological functionality \cite{Holzapfel2007,Ciarletta2016,Balbi2014,Fernandez-Sanchez2015,Stylianopoulos2012}.

In the context of epithelial tissues, there has been recent progress understanding the mechanics of the epithelia, and tools have now been developed to control mechanical forces in these tissues. 
The Madin-Darby canine kidney (MDCK) cell line, for example, has been widely used to study epithelial cells. 
It has been observed, for instance, that dKO mutant (double-knockout of scaffolding proteins ZO-1 and ZO-2) MDCK cells exhibit apical constriction \cite{Elia2009,Imai2015,Wells2013}.
By micropatterning wild-type and mutant cells together, one can therefore spatially pattern stress gradients on an MDCK monolayer's apical side, while leaving the basal side unaffected. 
Moreover, optogenetic tools have been developed to dynamically control apical constriction in mammalian epithelial cells~\cite{Martnez-Ara2022}. 
Shape programming living surfaces with nematic patterning has been achieved by growing orientation patterned fibroblast layers on soft substrates~\cite{Guillamat_2025}. 

We here focus on a disk with finite thickness, where incompatible in-plane growth patterns are programmed across the sheet’s thickness (Fig.~\ref{Fig1}(a)). 
Such a disk can therefore be thought of as a model of either epithelial tissue with in-plane patterns of apical constrictions or shape programmable gel with in-plane patterns of cross-linking.
Our goal is to understand how intrinsic curvatures and area mismatch interact and compete to reach non-trivial energetically stable shapes, potentially with significant residual strains post relaxation. 
Since the aspect ratio of the sheet influences the relative contributions of bending and stretching energies, we are particularly interested in how both the effective thickness and the growth gradient affect the resulting configuration.  
From a practical standpoint, our framework could be useful for tissue engineering or as a way to manipulate curvature in a cell sheet, allowing for the investigation of how cells respond to changes in geometry and mechanical stress.
The simulations are carried out based on a spring lattice model (\ref{S: simulation}) and solved via the \texttt{TopoSPAM}~\cite{Singh2025} framework in \texttt{OpenFPM}~\cite{INCARDONA2019}. The resulting curvature profiles are subsequently quantified using \texttt{pymeshlab}~\cite{pymeshlab}. 

We first present the results for uniform shrinkage of the apical surface to illustrate the competition between intrinsic and extrinsic curvature. We then examine cases where a passive surface competes with an anti-cone or a saddle shape, showing the different states these systems can adopt. Afterward, we demonstrate results from competitions involving a passive surface and positive curvatures. Next, we quantify the residual spontaneous strains in these competitions, presenting their spatial distributions as well as the influence of curvature parameters and aspect ratio on the strain within each layer. We then present selected examples from bilayers consisting of two simple patterned surfaces with distinct curvature signs and gradients.
Finally, we demonstrate generation of biological shapes, with examples of the \textit{Drosophila} ventral furrow and the multicusped dental epithelium, using a bilayer with simple, independent, incompatible strain patterns on each layer. 

\section{Swelling Pattern Competing with Passive Surface}
For a bilayer sheet when the top and bottom surfaces have different isotropic swelling gradient profiles, $\eta_a(r)$ and $\eta_b(r)$, respectively, and by assuming a linear gradient in deformation across the thickness, the volumetric swelling function can be written as 
\begin{equation}\label{3D_swell}
\eta(r,z) = \frac{1}{2}(\eta_b(r)+\eta_a(r)) + \frac{z}{h}(\eta_a(r)-\eta_b(r)),
\end{equation}
where \( z \in [-h/2, h/2] \) and \( r \in [0, R_0] \), with \( h \) denoting the thickness and \( R_0 \) the radius of the (reference) disk. The swelling factor \( \eta(r,z) \) describes the local isotropic in-plane expansion or contraction at point \( (r, z) \), where the apical and basal swelling factor $\eta_a(r)$ and $\eta_b(r)$ describes the local isotropic in-plane expansion or contraction on the apical and basal surface. 
We first study the deformation of a thick tissue sheet where one surface (apical) undergoes programmed homogeneous, isotropic, in-plane growth while the opposite surface (basal) remains passive, therefore, $\eta_b(r) = 1$. 
In the following, we later explore swelling patterns programmed in the apical surface with simple geometries, saddle, anti-cone, spherical cap, and cone, while varying the aspect ratio $\alpha$, defined as the diameter to thickness ratio, and the curvature associated with each metric.

\subsection{Uniform apical swelling or contraction}
\begin{figure*} 
\centering
\includegraphics[width=\textwidth]{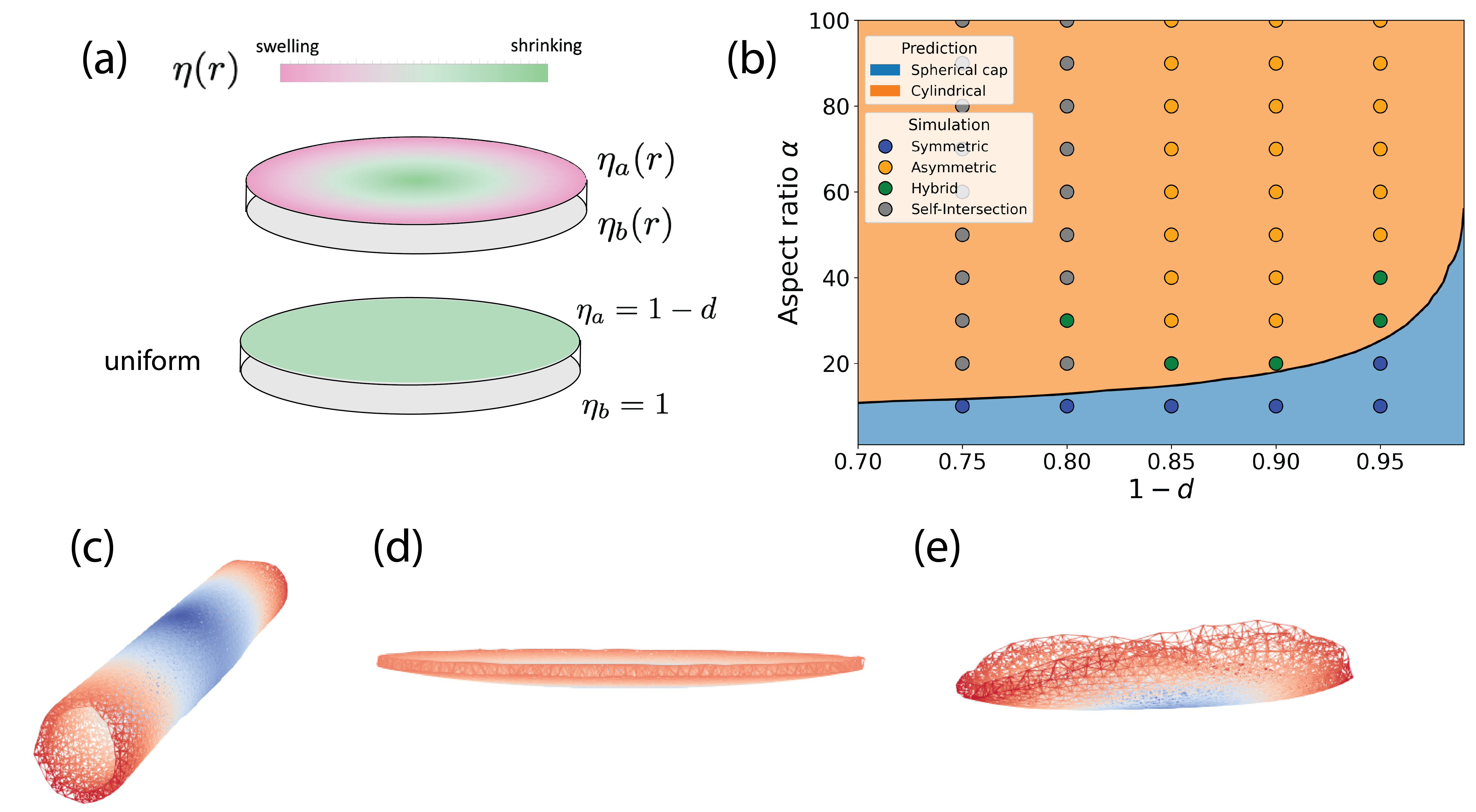}
\caption{(a) Schematic of a swelling pattern of a bilayer, where $\eta_a(r)$ and $\eta_b(r)$ denote the swelling functions on the apical and basal surfaces respectively. (b) Phase diagram of the equilibrated configurations under uniform shrinkage applied to the apical surface. The background colors are given by analytical predictions, while each point represents a simulation result for the corresponding parameters. Self-intersection denotes numerical breakdown, where local spring crossings occur due to incompatible deformations, leading to crumpling. (c)-(e) Representative shapes corresponding to the states in (b), colored blue-to-red from the center of the initial reference disk to its edge for ease of visualization. (c) A cylindrical shape for $\alpha = 60$ and $d = 0.15$. (d) A spherical cap for $\alpha = 25$ and $d = 0.01$. (e) An hybrid state near the phase boundary for $\alpha = 20$ and $d = 0.1$.}\label{Fig1}
\end{figure*}

Before analyzing more complex cases, we first consider a uniformly shrinking apical surface, with \( \eta_a(r) = 1 - d \). This scenario has been previously studied in PVS elastomer systems~\cite{Pezzulla_2016}, where the resulting shape varies between dome-like and cylindrical depending on the amount of swelling mismatch.
Here, we explore how both the aspect ratio and shrinkage magnitude affect the final shape, assuming it takes either a spherical cap or a cylindrical profile. Using analytical energy expressions and spring-lattice simulations, we construct a phase diagram (Fig.~\ref{Fig1}(b)). The elastic energies for the cylinder and spherical cap are given by:
\begin{widetext}
\begin{equation}
E_\text{c} = \frac{EV}{24(1-\nu^2)}\frac{1}{\gamma_c^2\alpha^2}
+
\frac{EV}{48}\left(d^2 + \left(\frac{2-d}{\gamma_c\alpha}-d\right)^2\right)^2,
\end{equation}
\begin{equation}
\begin{aligned}
E_s &=
\frac{EV}{1-\nu^2}\frac{1-\cos\phi_0}{3\alpha^2}
+
\frac{EV}{2}\Big[
2\left(1-d+\frac{d^2}{3}\right)
+
\left((2-d)\gamma_{s} + \frac{d}{3\alpha}\right)
\left(\frac{2(\cos\phi_0-1)}{\phi_0}-\phi_0\right)
\\
&\qquad\qquad+
\left(\gamma_{s}^2 + \frac{1}{3\alpha^2}\right)
\left(\phi_0^2 - \mathrm{Ci}(2\phi_0) + \log(2\phi_0) + \gamma\right)
\Big],
\end{aligned}
\end{equation}
\end{widetext}
where $E$ is the Young's modulus, $\nu$ is the Poisson's ratio, and $V$ is the volume of the sheet. 
$\text{Ci}(\cdot)$ is the cosine integral and $\gamma$ is Euler's constant, $\gamma_c$ and $\gamma_s$ are the dimensionless radius of curvature of the cylinder and the sphere, normalized by \(R_0\), respectively, and $\phi_0$ is an angle that determines the spherical cap area for a fixed radius of curvature $\gamma_s$ (Fig.~\ref{fig: spherica_cap_schematic}).
We minimize each energy numerically with respect to \( R_\text{c} \) or \( R_\text{s} \), and the shape with lower energy is considered energetically favorable. Detailed derivations are provided in the supplementary material (\ref{S.1}).

Since both layers are initially flat and the active swelling on the apical surface is uniform and induces no intrinsic curvature on its own, each of the two layers will exhibit an intrinsic curvature-free ground state. The global swelling mismatch between the surfaces, on the other hand, induces a preferred extrinsic curvature. The phase diagram shows that when the sheet is relatively thick, extrinsic effects dominate, favoring a dome-like shape (Fig.~\ref{Fig1}(d)) with uniform non-zero Gaussian curvature.
In contrast, when the sheet is thin or the swelling difference is large, intrinsic curvature becomes dominant with a developable surface. The sheet curls to accommodate area mismatch, with each layer remaining intrinsically flat, i.e. with zero Gaussian curvature (Fig.~\ref{Fig1}(c)).
Near the phase boundary, simulations reveal hybrid shapes: a relatively flat center with positively curved edges, especially along the longer axis (Fig.~\ref{Fig1}(f)). In this intermediate regime, the system minimizes energy by combining intrinsic and extrinsic curvature effects.

\subsection{Negative Curvature programmed on the apical surface}
Moving beyond uniform swelling or contraction, we now introduce patterns of extensile or contractile activity on the apical surface, while keeping the basal surface passive. We first explore two different negative intrinsic curvature patterns competing with a passive surface: one with uniform curvature (i.e., a saddle) and one with localized curvature at the center (an anti-cone). The isotropic swelling functions that program a saddle and an anti-cone shape are given by
\begin{align}
\eta_\text{s-}(r) &= \frac{2a\gamma_\text{K}R_0}{a^2 - r^2}, \label{eq:saddle_swelling} \\
\eta_\text{a-}(r) &= \left(\frac{r}{R}\right)^{1 - \sin\phi}, \label{eq:anticone_swelling}
\end{align}
respectively, over the radial domain \( r \in [0, R_0] \). Derivation of the expressions are provided in the supplementary material \ref{S:swelling_functions}. The curvature parameter, \( \gamma_\text{K} \), is the dimensionless radius of curvature of the saddle, resulting in a local Gaussian curvature of \( -(\gamma_\text{K} R_0)^{-2} \). For the anti-cone, the curvature parameter \( \phi_0 \) refers to the angular surplus at the center, producing a total Gaussian curvature of \( -2\pi(1 - \sin\phi) \). The constants \( a \) and \( R \) are size scaling parameters for the saddle and anti-cone, respectively, and are chosen such that the total in-plane area of the programmed apical surface matches that of the passive basal surface.
Since a mismatch in total surface area of the apical and basal layers can induce a global curvature across the entire sheet, as shown in the previous section, the area of the deformed apical surface is constrained to be preserved to ensure that the effects arising comes purely from curvature differences and the local competition.
Illustrations of the two swelling functions at different curvatures are shown in Fig.~\ref{Fig: saddle}(b)  and Fig.~\ref{Fig: anticone}(b).

\subsubsection{Saddle}
\begin{figure*} 
\centering
\includegraphics[width=\textwidth]{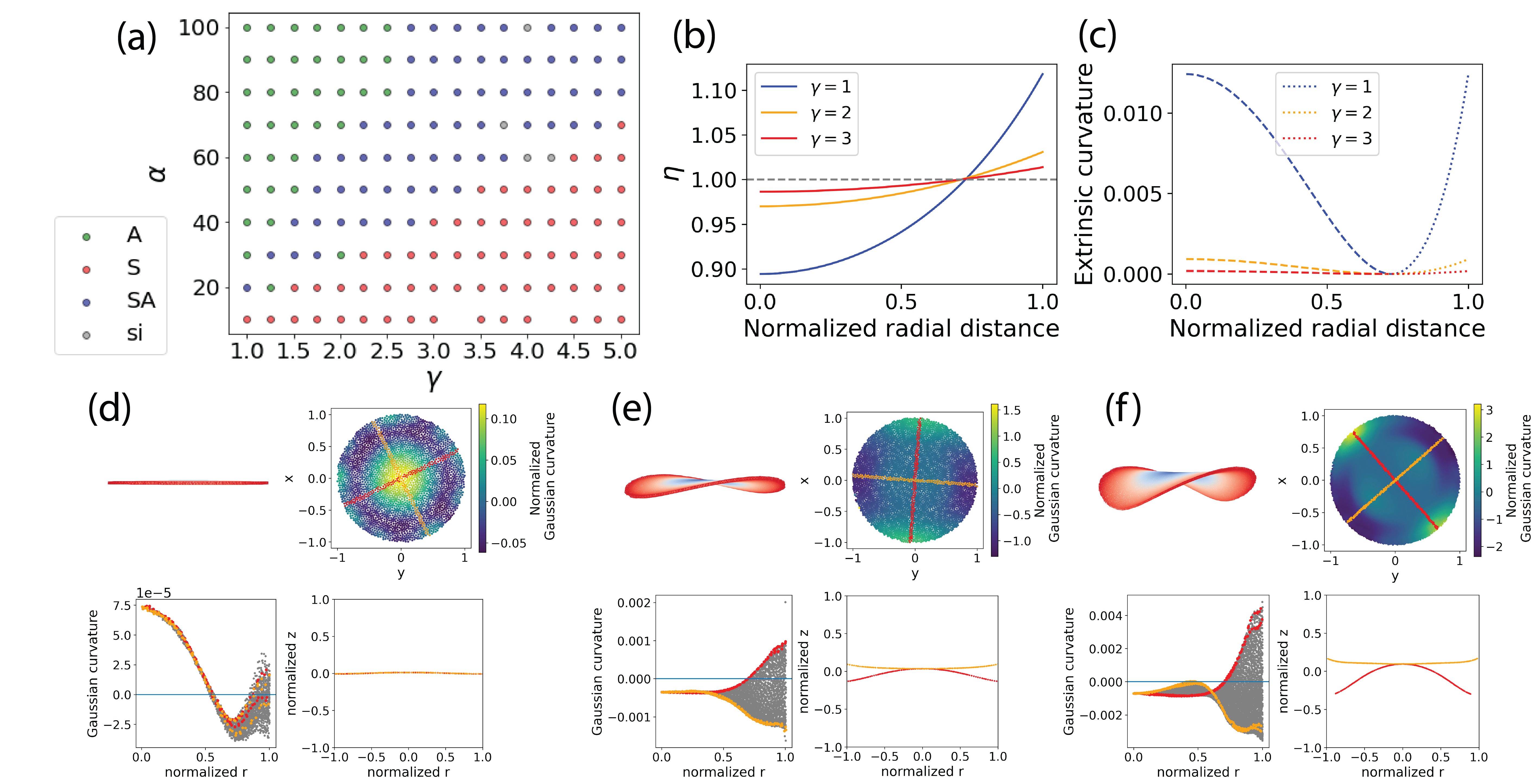}
\caption{Simulation results for competition between a passive layer and one with a saddle-shaped target curvature. (a) Phase diagram. States are characterized by the symmetry in Gaussian curvature: S - symmetric, SA - symmetric near center and asymmetric after some radius, A - asymmetric from center. State si refers to the states that have self-intersections in the final configuration, and have not been further analyzed.
(b) Swelling function $\eta_\text{s-}(r)$ corresponding to different target radius of curvature of $\gamma_\text{K}R_0$.
(c) Predicted extrinsic curvature induced by the swelling profile on the apical surface. The dashed line and dotted line refers to a upward and downward bending respectively. 
(d–f) Representative configurations for each distinct state. For each configuration:
Top left: Snapshot of the 3D configuration, with a blue-to-red color gradient denoting radial distance for visualization.
Top right: Normalized Gaussian curvature profile mapped onto the initial flat configuration. It is normalized by a factor $(\gamma_K R_0)^2$. Red and orange lines denote the principal axes of symmetry breaking.
Bottom left: Gaussian curvature as a function of radial distance, with points along the principal axes highlighted in red and orange.
Bottom right: Normalized height $z$ plotted against normalized radial position $r$ for the points along the two principal axes.
}
\label{Fig: saddle}
\end{figure*}

Although the total areas are equal, the local swelling and deswelling induce local area mismatches with the passive sheet, giving rise to localized preferred extrinsic curvatures (Fig.~\ref{Fig: saddle})(c). As a result, the system exhibits a three-way competition among the flat and negatively curved intrinsic curvatures of the two sheets, respectively, and the extrinsic curvatures induced by local area mismatches.
Three distinct shape states emerge, depending on the aspect ratio and magnitude of the programmed intrinsic curvature. In some cases the final shape spontaneously breaks rotational symmetry. For these cases, we identify the symmetry-breaking axes and classify the resulting states by comparing the curvature along these principal directions.
In the regime of high thickness and small intrinsic curvature, the curvature profiles remain symmetric (state S, Fig.~\ref{Fig: saddle}(d)), and the Gaussian curvature stays positive up to a boundary penetration depth whose length scale is set by the thickness. This is due to the dominance of extrinsic curvature in this parameter regime.
As the intrinsic curvature or aspect ratio increases, the curvature at the center transitions to negative values  (state SA, Fig.~\ref{Fig: saddle}(e)). This shift indicates that maintaining a symmetric state becomes energetically unfavorable due to the stretch energy, and the intrinsic curvature begins to dominate. However, the curvature profiles along the long and short axes begin to diverge beyond a certain radial distance.
In the opposite extreme, the curvatures break symmetry with one another already from the center (state A, Fig.~\ref{Fig: saddle}(f)). Along the long axis (red line), the curvature remains nearly constant up to a point before increasing. In contrast, along the short axis, the curvature initially increases but then drops to a negative value. This behavior may be attributed to a higher-order buckling effect due to the larger local mismatch.

\subsubsection{Anti-cone}
The primary difference in programming an anti-cone metric on the apical surface versus a saddle metric is that the anti-cone is (metrically) flat everywhere except at the center. Due to the finite thickness of the sheet, the negative curvature is not confined exactly at a point but spread out up to a length related to the thickness. Therefore, the curvature induced by the swelling in the apical surface has a radial decay in Gaussian curvature from the center. Near the center, we have the competition of three curvatures, while the metric of both surfaces is flat away from the center and hence no negative curvature is expected to come into play here. 

\begin{figure*} 
\centering
\includegraphics[width=\textwidth]{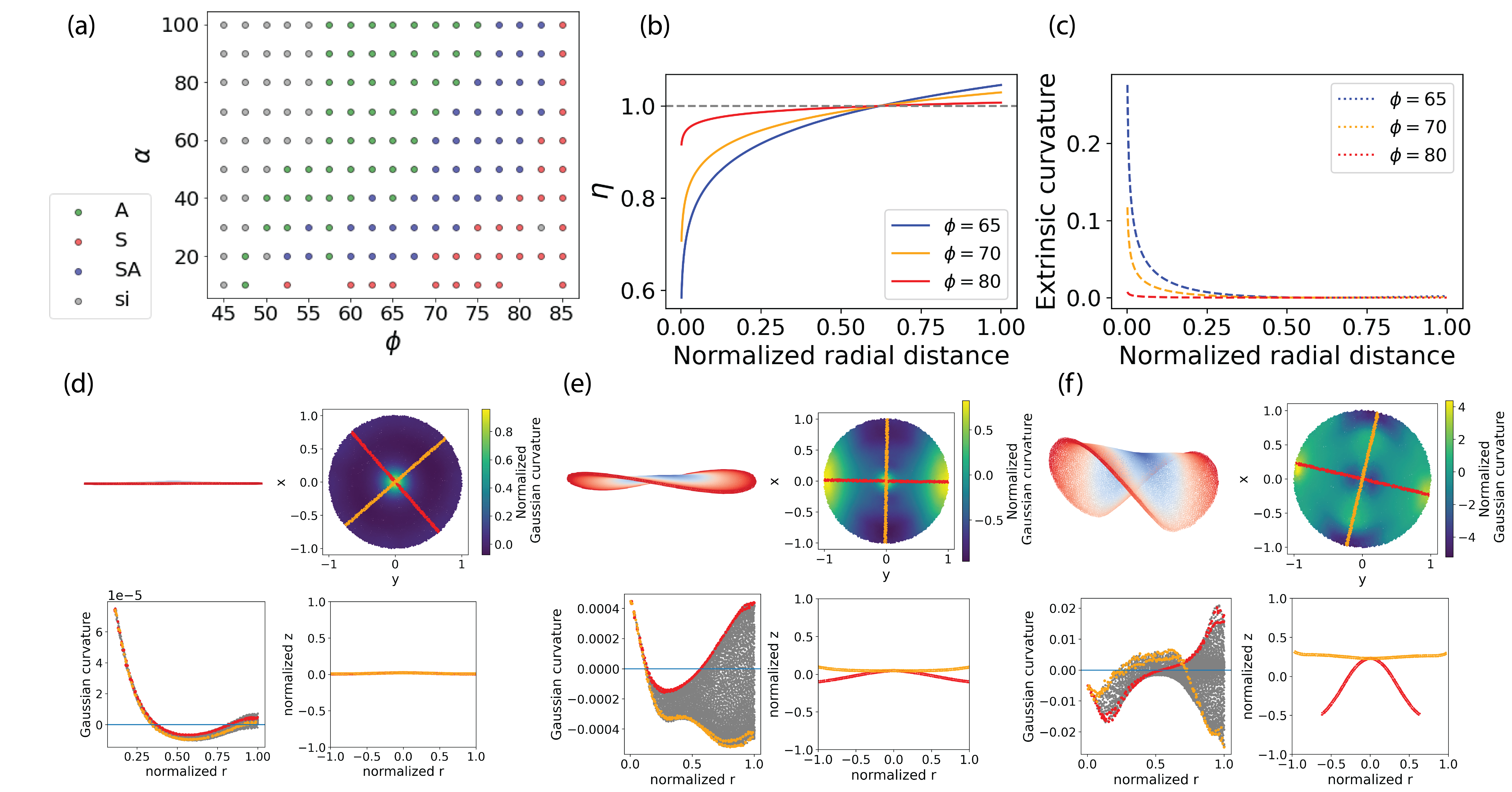}
\caption{
Simulation results for competition between a passive layer and one with an anti-cone target curvature. (a) Phase diagram. States are characterized by the symmetry in Gaussian curvature: S - symmetric, SA - symmetric near center and asymmetric after some radius, A - asymmetric from center.
(b) Swelling function $\eta_\text{a-}(r)$ corresponding to different curvature parameters $\phi$.
(c) Predicted extrinsic curvature induced by the swelling profile on the apical surface. The dashed line and dotted line refers to a upward and downward bending respectively. 
(d–f) Representative configurations for each distinct state. For each configuration: Top left: Snapshot of the 3D configuration, with a blue-to-red color gradient denoting radial distance for visualization. Top right: Normalized Gaussian curvature profile mapped onto the initial flat configuration, normalized by $R_0^2/K$. %check the normalization
Red and orange lines denote the principal axes of symmetry breaking. Bottom left: Gaussian curvature as a function of radial distance, with points along the principal axes highlighted in red and orange. Bottom right: Normalized height $z$ plotted against normalized radial position $r$ for the points along the two principal axes.
}
\label{Fig: anticone}
\end{figure*}

We again identify three different shape states, each distinguished by the behavior of Gaussian curvature along the two principal axes (Fig.~\ref{Fig: anticone}(a)).  In the case of relatively thick sheets and low prescribed curvature, the shape remains radially symmetric, and the Gaussian curvature decays very rapidly from the center as compared to the saddle case (state S, Fig.~\ref{Fig: anticone}(d)). 
In the intermediate regime, due to the high magnitude of area mismatch at the center, it retains positive curvature, in contrast to the saddle pattern where the central curvature is typically negative and even (state SA, Fig.~\ref{Fig: anticone}(e)). 
Although the curvature initially remains radially symmetric, it gradually transitions into an asymmetric distribution away from the center. The extrinsic curvature emerges towards the boundary of the disk, resulting in positive curvature near the red axis, and reducing the amount of negative curvature at the ends of the orange axis.
At the opposite extreme, symmetry is broken already moving away from the center (state A, Fig.~\ref{Fig: anticone}(f)).  and a complex shape develops reminiscent of those found in anti-clastic suppression scenarios, with an oppositely curved domain close to a free boundary~\cite{Warner_2010}.
Note that, even though there is no local source for negative intrinsic curvature away from the center, we consistently observe it in these competitions (Fig.~Fig.~\ref{Fig: anticone}(d-f)).
This could be accounted for by the spread of the localized curvature from the center due to the thickness, as has been described in monolayer shape-programmed systems~\cite{Modes2011-1} or lead to by a geometric incompatibility that will be more clear in the context of competing positive curvatures and that we will accordingly discuss in the next section.

\subsection{Positive Curvature programmed on the apical surface}

We now consider cases where the apical surface is programmed with either uniform or localized positive intrinsic curvature, corresponding to a spherical cap or a conical surface, respectively. The isotropic swelling function for these geometries is given by
\begin{align}
\eta_\text{s+}(r) &= \frac{2a\gamma_\text{K}R_0}{a^2 + r^2}, \label{eq:spherical_cap_swelling} \\
\eta_\text{c+}(r) &= \left(\frac{r}{R}\right)^{\sin\phi - 1}, \label{eq:cone_swelling}
\end{align}
respectively (\ref{S:swelling_functions}), defined over the radial domain \( r \in (0, R_0) \) of the reference disk. The parameters \( \gamma_\text{K} \), \( \phi_0 \), \( a \), and \( R \) are defined analogously to those in Equations~\eqref{eq:saddle_swelling} and~\eqref{eq:anticone_swelling}. \( a \) and \( R \) are chosen such that the total target in-plane area of the apical layer matches that of the passive basal surface. These swelling profiles with \( \gamma_\text{K} \) and \( \phi \) encode the same magnitude of Gaussian curvature as their negative-curvature counterparts, but with a positive sign.

\begin{figure*} 
\centering
\includegraphics[width=\textwidth]{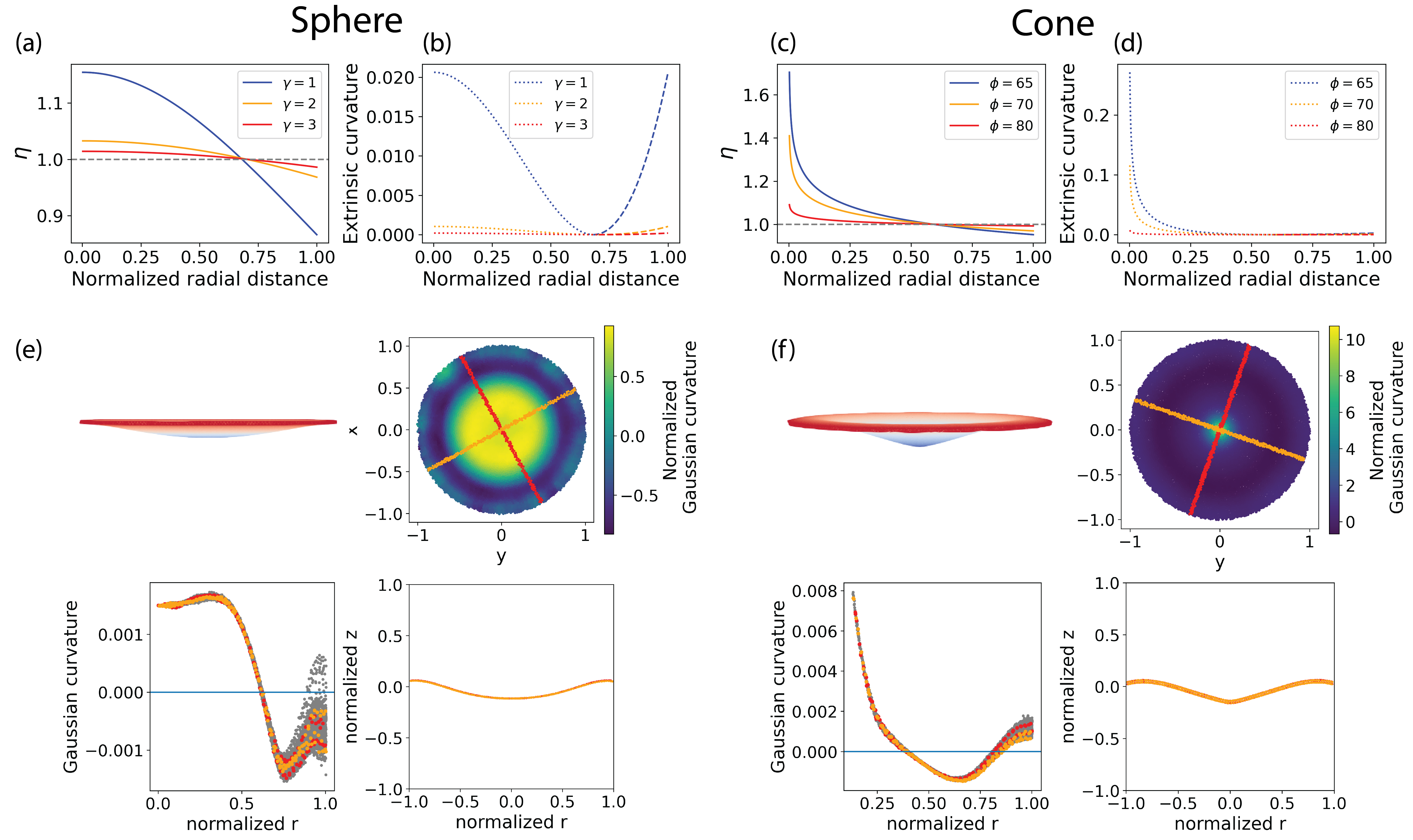}
\caption{
(a,c) Swelling functions of sphere and cone patterns.
(b,d) Predicted extrinsic curvature induced by the swelling profile on the apical surface. The dashed line and dotted line refers to a upward and downward bending respectively. 
(e–f) Top left: Snapshot of the 3D configuration, with a blue-to-red color gradient denoting radial distance for visualization. Top right: Normalized Gaussian curvature profile mapped onto the initial flat configuration, normalized by $(\gamma_K R_0)^2$. Red and orange lines denote the principal axes of symmetry breaking. Bottom left: Gaussian curvature as a function of radial distance, with points along the principal axes highlighted in red and orange. Bottom right: Normalized height $z$ plotted against normalized radial position $r$ for the points along the two principal axes.}
\label{Fig: sphere_cone}

\end{figure*}

As before, the gradient in swelling induces extrinsic curvature (Fig.~\ref{Fig: sphere_cone}(b,d)), and both the intrinsic and extrinsic curvatures are of the same sign. Because the circumferential arc length is less than that of a circle with the same in-material radius in the apical surface, azimuthal compression is not strong enough to break rotational symmetry. This leads to configurations that remain radially symmetric, with a similar curvature gradient and variation in curvature magnitude at the boundary, depending on the parameters. The symmetric configurations from the saddle or anti-cone also have a similar curvature profile.  

Interestingly, despite the absence of any hyperbolic contribution from the area mismatch or surface metric, the resulting shapes consistently exhibit regions of negative Gaussian curvature (Fig.~\ref{Fig: sphere_cone}(e-f) bottom left). This effect does not originate from the intrinsic or extrinsic curvature alone but rather from a geometric incompatibility. Although bending up and down is not defined in a purely 2D metric, extrinsic curvature has an oriented out-of-plane direction. Since the apical layer contains both shrinking and swelling regions, it promotes bending in opposite directions near the center and the edge. This must involve a saddle-like transition zone, enforced by the geometric constraints of embedding in Euclidean space, and subsequently drives the establishment of the observed negative curvature.

\subsection{Universal Scaling of Normalized Curvature Response} 
\begin{figure*}
\centering
\includegraphics[width=0.6\textwidth]{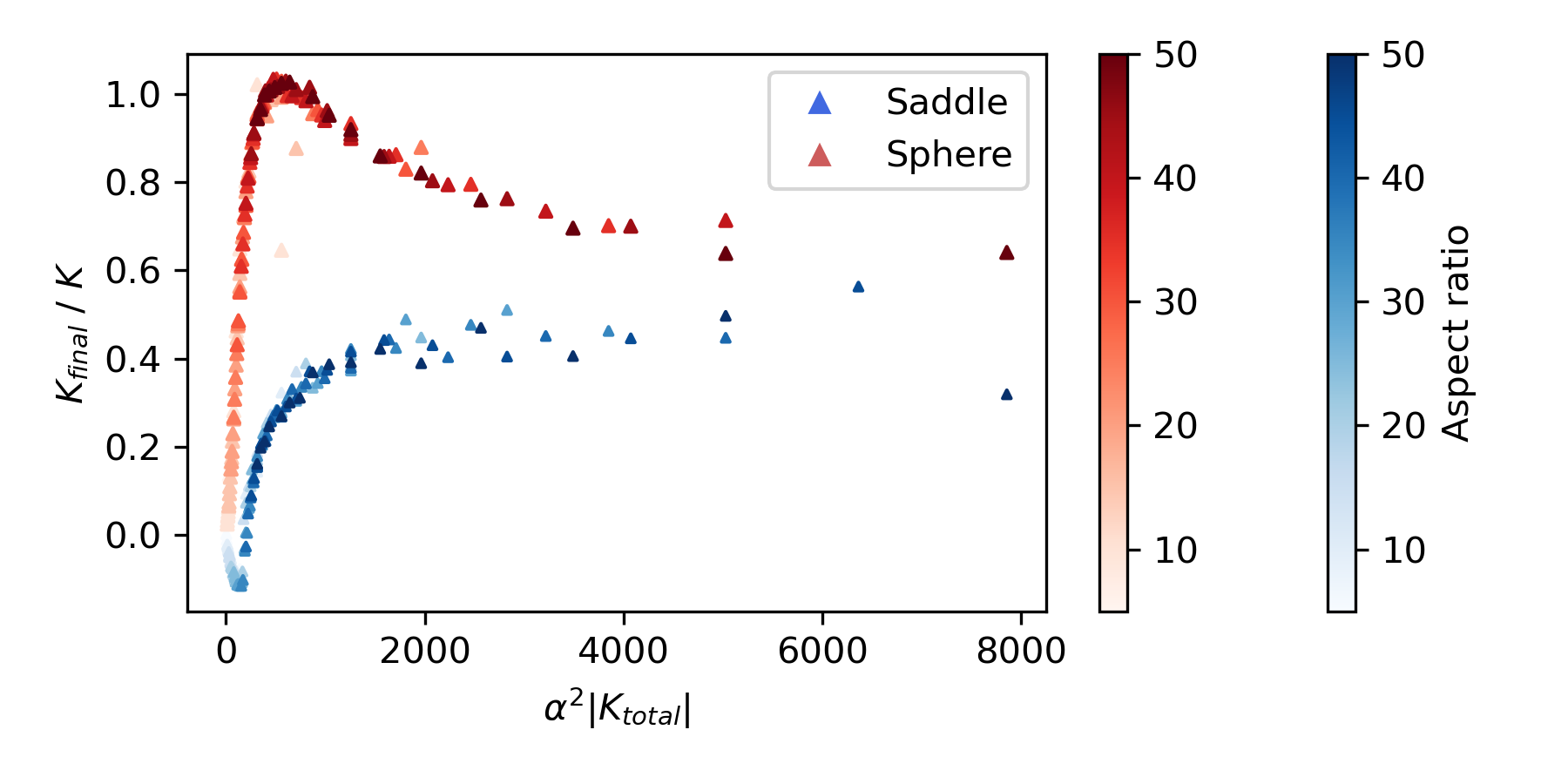}
\caption{Universal scaling behavior of the normalized Gaussian curvature located at the center of the disk.
Measured Gaussian curvature in the final configuration, normalized by the target curvature, is plotted against $\alpha^2|K_{\text{total}}|$ for both spherical cap and saddle patterns. 
Here, $K_{\text{total}}$ represents the target integrated Gaussian curvature over the pattern surface. 
Red and blue points denote the results for the spherical cap and saddle target patterns, respectively.
}
\label{Fig: curvature_collapse}
\end{figure*}

As seen in Fig.~\ref{Fig: sphere_cone} and Fig.~\ref{Fig: saddle}, the local Gaussian curvature at the center of the shape converges to a definite value.
Interestingly, when plotting this curvature normalized by the target local curvature of the active surface, we observe evidence of universal scaling behavior when this normalized value is plotted against the dimensionless quantity $\alpha^2 K_\text{total}$, where $K_\text{total}$ is the total curvature programmed into the active surface.
This universal scaling behavior is typified by curve collapse of normalized curvature versus $\alpha^2 K_\text{total}$ (Fig.~\ref{Fig: curvature_collapse}).
In the case of the competition with the sphere, the normalized curvature starts from zero, corresponding to the situation where the curvature of the top surface is very weak and the surface is thick, so the shape remains in its original flat configuration. 
The normalized curvature then increases until it reaches $1$, meaning that the curvature matches the target curvature. 
Here, where the local curvature at the center achieves the target curvature suggests a critical $\alpha^2K_\text{total}$ that minimizes any incompatibility between the extrinsic and intrinsic curvatures at the center of the shape.
Below this critical $\alpha^2 K_\text{total}$, we can interpret the failure of the center curvature to reach its target as arising from a thickness-induced incompatibility with the local extrinsic curvature. Above this value of $\alpha^2 K_\text{total}$, the normalized curvature drops again, possibly due to the tighter coupling for thinner sheets between the preferred intrinsic curvature of the two sheets.
For the saddle case, the normalized curvature first drops below $0$, representing the $S$ state in which the curvature at the center is positive. 
After that, the curvature increases again and saturates around $0.4$.
Unlike the case of the spherical cap, the curvature ratio here never reaches 1 because the positive extrinsic curvature is always at odds with the target (intrinsic) negative curvature.

\subsection{Residual strains}
We now consider the residual spontaneous strains that remain unfulfilled after deformation on each layer and their combination. 
In our spring lattice simulations, we evaluate these residual strains based on the final, equilibrated length of each spring and the target length of each spring as follows:
The in-plane strains on spring $i$ in the apical and basal layers are denoted $\epsilon_{i,a}$ and $\epsilon_{i,b}$, defined as $\epsilon_{i} = (l_{i,\text{current}} - l_{i,\text{target}})/l_{i,\text{target}}$. 
The cross-layer spring strains, which we denote $\epsilon_{i,c}$, are defined similarly.
The average strain over the apical surface is $\langle \epsilon_{a} \rangle = \frac{1}{N}\sum_{i=0}^N \epsilon_{i,a}$, representing net expansion or contraction. 
The average absolute strain, $\langle |\epsilon_{a}| \rangle = \frac{1}{N}\sum_{i=0}^N |\epsilon_{i,a}|$, measures the total strain deviation from a fully relaxed sheet and is related (by a square root) to the stored effective energy after relaxation of the imposed spontaneous strains. These quantities are similarly defined for the basal surface.

\begin{figure*} 
\centering
\includegraphics[width=0.9\textwidth]{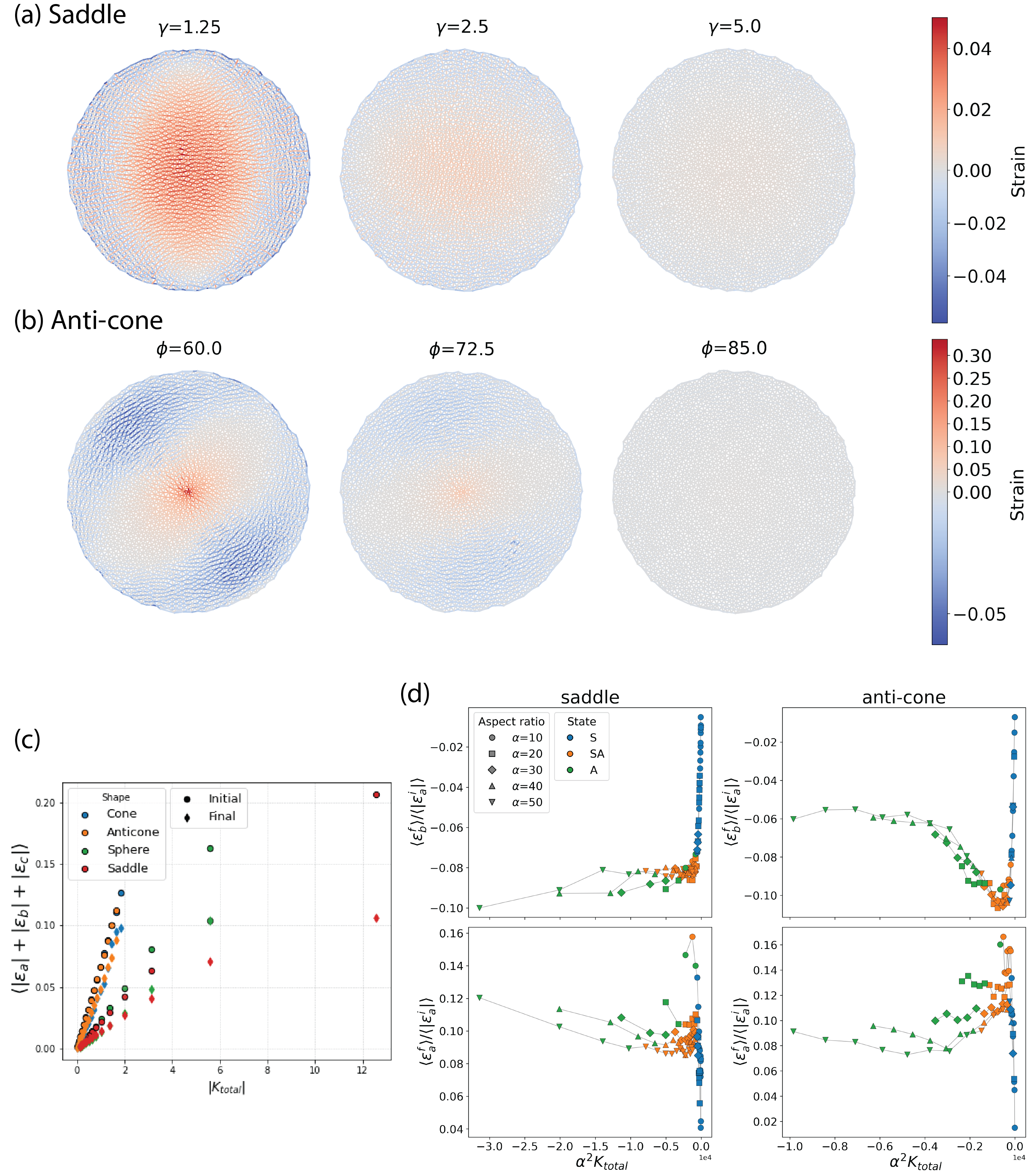}
\caption{(a-b) Residual strain spatial profiles of the three distinct states for saddle and anti-cone patterns competing with a passive surface. The strain strength is normalized within each shape. (c) Average absolute strain over the entire system plotted against the target total curvature of the active surface  . The dependence on the aspect ratio is removed by taking the mean values of the strains over the aspect ratios. (d) The curve collapse of average final strain normalized by the average absolute initial strain, plotted against $\alpha^2 K$ for shapes from saddle and anti-cone competitions, with the three states colored differently, each occupying its own region of the collapsed curve.}
\label{Fig: strain}

\end{figure*}

In the initial state, the swelling region is instantaneously compressed (negative strain) and the shrinking region is instantaneously stretched (positive strain).
The initial average absolute strain, $\langle |\epsilon_{a}^i|+|\epsilon_{b}^i|+|\epsilon_{c}^i| \rangle$, scales linearly with the total curvature of the target shape, but with different target shapes exhibiting different slopes (Fig.~\ref{Fig: strain}(c)). There is little to no apparent dependence on the aspect ratio (\ref{S.strain_R0}). 
Meanwhile, the final average absolute strain, $\langle |\epsilon_a^f| + |\epsilon_b^f| +  |\epsilon_c^f| \rangle$, also scales linearly with the total curvature, as in the initial state.
Here however, light dependence on the aspect ratio appears, particularly for thicker bilayers as shown in (\ref{S.strain_R0}). 
By comparing lines associated to the final average absolute strain to those associated to the initial average absolute strain, it is surprising to note that the total amount of relaxation achieved during the establishment of the shape is relatively small. 

It is interesting to investigate the average residual strain of each surface for each negatively curved shape, normalized by the average absolute initial strain on an index surface, which we here choose to be the apical surface (similar analysis for the positively curved shapes are in \ref{S: normalized_strain}.
One can think of this quantity as a metric that reflects the relative amount of remaining net stretch or compression after each shape relaxes.
Note that when this metric is plotted against $\alpha^2 K_\text{total}$, the same quantity as in Fig~\ref{Fig: bio_shapes}, curves associated to different aspect ratios of the bilayer all collapse together, again indicating a universal scaling behavior.
Interestingly, the same curve collapse with the same scaling occurs for each different shape although in each case the form of the collapsed curve is different (\ref{S: curve_collapse}).
Furthermore, the classes of behavior identified in Fig.~\ref{Fig: saddle} and \ref{Fig: anticone} are cleanly reproduced on each collapsed curve, with different sections of the curve corresponding to different phases (Fig.\ref{Fig: strain}(d)).

Finally, we examine the spatial distribution of the residual strains for the negatively curved shapes (positively curved shapes in the supplementary \ref{S: distribution}). For the saddle, representative residual maps taken from each phase display relatively minor deviation from circular symmetry even in the shape phases that strongly break symmetry. This likely indicates that when strong breaking of shape symmetry occurs, it is a consequence of incompatibilities associated to the spatial embedding rather than the direct competition of the in-plane metrics of the two sheets. 
Similarly, for the anti-cone, there is even less apparent deviation from circular symmetry while at the same time significantly larger residual strains develop near the anti-cone tip, creating larger localized energy storage.

\subsection{Competition between patterns}
Moving beyond one active layer competing against an inert surface, we now turn to competition between two active surfaces, each with distinct curvature profiles, namely combinations of cone, anti-cone, sphere, and saddle. 
Our goal here is not to analyze the full parameter space of all possible shapes, which becomes very complex, but rather to highlight representative and interesting outcomes for each pair of patterns. The simulations are performed with an aspect ratio of $\alpha = 60$ under varying curvature strengths. Within the prescribed curvature range, this aspect ratio is not one where extrinsic curvature alone dominates, allowing us to explore an interesting region of the state space. 

\begin{figure*} [t]
\centering
\includegraphics[width=0.8\textwidth]{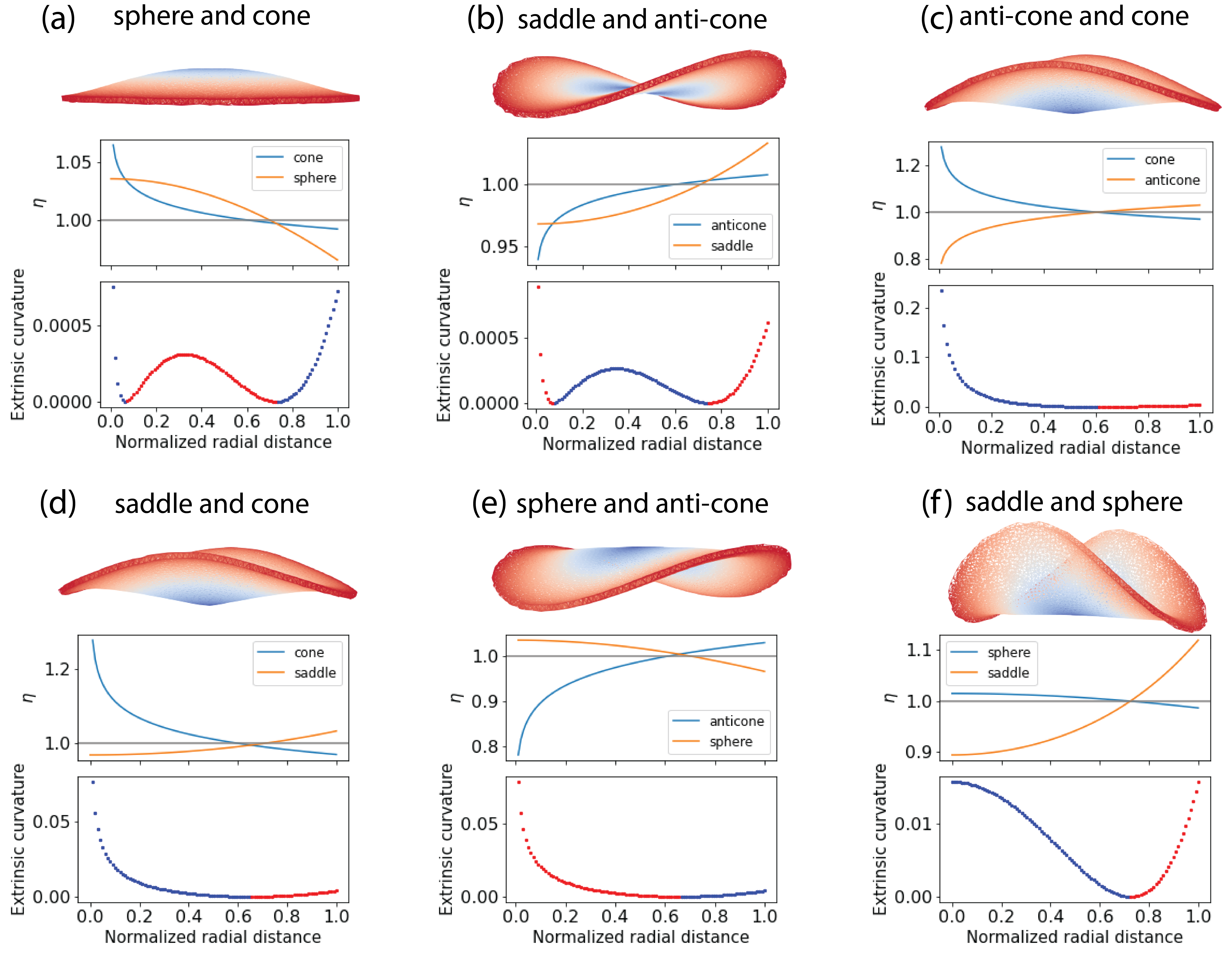}
\caption{Representative results of competition with two active patterns with distinct curvature profiles, colored blue-to-red from the center of the initial reference disk to its edge for ease of visualization. The sequence of the two geometries in each title refers to the apical and basal patterns respectively. The plots below each shapes are the swelling function of each shape for the specific parameter set and the expected extrinsic curvature approximated from the difference in swelling. Blue and red points in the extrinsic curvature plots distinguish between opposite bending directions.} 
\label{Fig: competition}

\end{figure*}
For patterns with the same sign of intrinsic curvature, i.e., a sphere competing with a cone, or a saddle with an anti-cone, we find that for certain prescribed curvatures, the difference in swelling functions changes sign twice between $r = 0$ and $r = R_0$. This implies that the bending direction determined by extrinsic curvature is inverted twice (Fig.~\ref{Fig: competition}(a-b)). In the case of a sphere-cone mismatch, since both shapes preserve rotational symmetry, the resulting configuration also remains symmetric. For parameters where the bending direction switches twice, we observe a configuration with a cone tip pointing downward at the center, and the outer region behaves like a spherical cap bulging upward, and the rim curls up slightly (Fig.~\ref{Fig: competition}(a)). 
In contrast, for the negative-curvature case, no such inversion is observed despite having the same extrinsic curvature trend due to their symmetry breaking (Fig.~\ref{Fig: competition}(b)). 

When surfaces with curvatures of opposite signs compete, the outcome depends on the relative curvature strengths. 
The final configuration tends to resemble more closely the surface with higher curvature. In the competition between a cone and an anti-cone, or between a cone and a saddle, we find that if the cone curvature is slightly smaller than the saddle’s, or if the cone and anti-cone curvatures have equal magnitude, a cone tip forms at the center, accompanied by global symmetry breaking (Fig.~\ref{Fig: competition}(c-d)). A similar effect occurs for a sphere competing with an anti-cone, that the center forms a symmetry-breaking spherical cap (Fig.~\ref{Fig: competition}(e)). For the case of a sphere versus a saddle, when the saddle curvature is much stronger, interestingly, the center remains flat rather than adopting the saddle profile (Fig.~\ref{Fig: competition}(f)). Simulation results of more curvature parameter pairs can be found in the supplementary \ref{S: competition}. 

\section{Programming biological shapes}
\begin{figure*} [t]
\centering
\includegraphics[width=\textwidth]{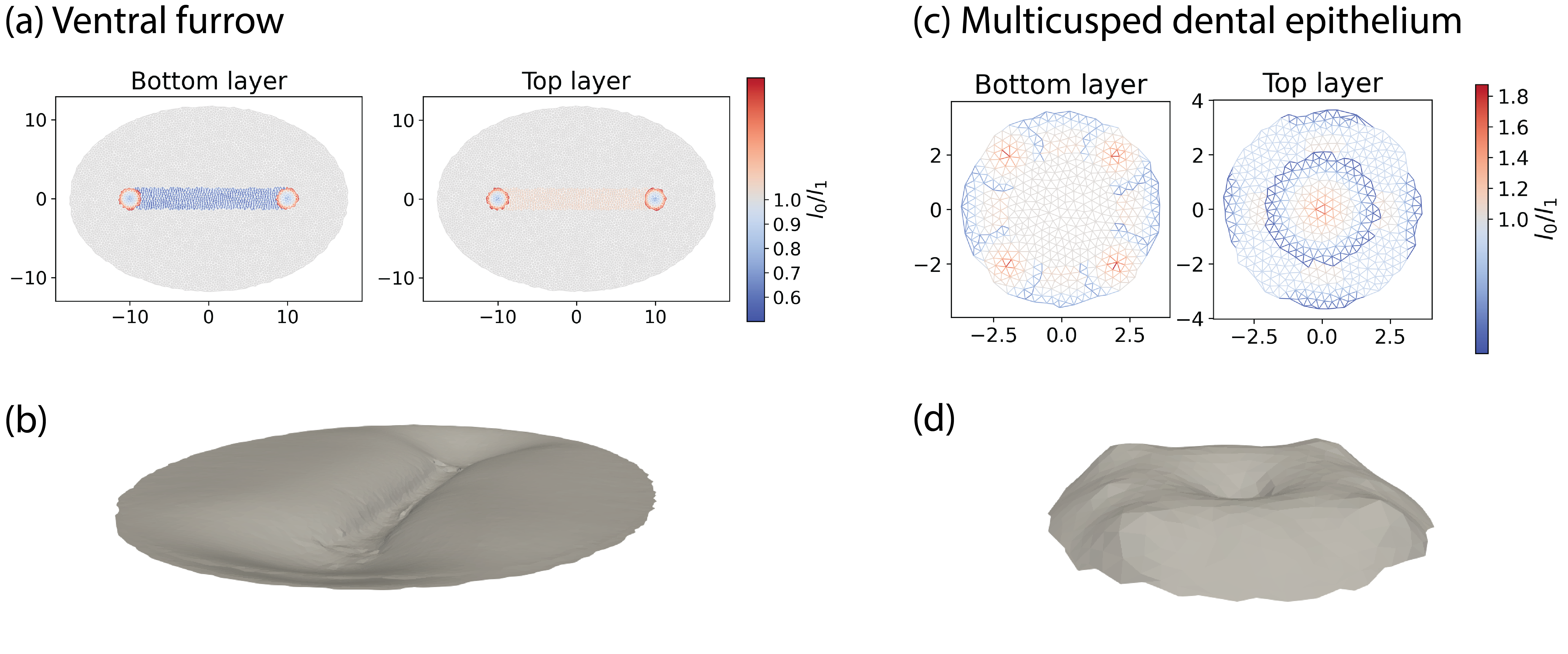}
\caption{Biologically inspired shape programming. (a,c) Spontaneous strain patterns to program the ventral furrow and multicusped dental epithelium shapes. (b,d) The force balanced final configurations of the two shapes. }
\label{Fig: bio_shapes}

\end{figure*}
Finally, we demonstrate that the morphologies of selected biological systems can be programmed using combinations of the above spontaneous strain patterns applied to a bilayer sheet. 
In particular, we reproduce two examples: the shape formation of the ventral furrow in \textit{Drosophila} embryogenesis and the multi-cusped dental epithelia.

In the first example, the ventral furrow in \textit{Drosophila}, we focus on a prominent feature that forms at an early stage of embryonic development. 
During this process, it is known that cells located on the ventral side of the embryo undergo pronounced, directional apical constriction, which rapidly drives a unidirectional inward folding into a tube-like structure~\cite{Sweeton1991,Conte2012,Leptin1991}. 
The literature often describes this process as being generated by a reduction in apical cell area, however, at the two ends of the furrow, this reduction alone would introduce substantial incompatible forces locally.
To demonstrate that the ventral furrow fold can be instead programmed in a way that could alleviate these high residual stresses, we impose the following strain patterns on an elliptical bilayer sheet with boundary points constrained to a constant $z$ position. 
A 50\% uniaxial contraction is applied to the apical surface over the furrow, and a small expansion is applied to the basal surface to maintain approximate cell-volume conservation. 
At the two ends of the furrow region, two anti-cone patterns are introduced to mediate the transition between the folded and flat regions and affect a significant reduction in local incompatible forces. 
The spontaneous strain pattern on the apical and basal surfaces, together with the resulting ventral furrow morphology, is shown in Fig.~\ref{Fig: bio_shapes} (a-b).

The second example is the shaping of the dental epithelium, upon which tooth formation relies. 
The epithelium initially undergoes budding, followed by an invagination that forms a cap-like structure, known as the bud-to-cap transition~\cite{Monteiro2018,Takigawa-Imamura2015}.
The cap stage subsequently develops into the bell stage, during which a single cusp or multiple cusps emerge, ultimately determining the final tooth shape. 
Here, we utilize spherical-cap and cone-shaped swelling gradients in a bilayer to program a four-cusped tooth morphology. 
The simulation starts with a flat bilayer with free boundary conditions. 
To generate the central cap shape, a cone pattern is applied to the top surface, while four smaller cone patterns are applied to the bottom surface. 
Because the direction of curvature is governed by the difference in swelling between the two layers, assigning the cone patterns to opposite sides induces bending in opposite directions robustly. 
On top of the cone-shaped patterns, a spherical-cap gradient with smaller curvature is applied to the part of the surface that do not carry a cone gradient to curve the overall shape as it is in the bell stage.
The spontaneous strain pattern on the apical and basal surfaces, together with the resulting dental epithelium, is shown in Fig.~\ref{Fig: bio_shapes} (c-d).

\section{Discussion}
We have demonstrated that materials composed of competing bilayers of active, isotropic, in-plane shape programmed materials are capable of exhibiting a wide array of non-trivial behaviors. We first investigated the case wherein the active surface shrinks uniformly, demonstrating a straightforward competition of intrinsic and extrinsic curvature, with both the aspect ratio and the degree of area-mismatch as parameters. Here, extrinsic curvature dominates in the thick limit and when the mismatch is low, and intrinsic curvature plays a more important role in the opposite limit. Next, we investigated a family of shapes that arise when a passive layer competes with an active layer programmed with positive or negative curvatures or different types. The competition with positive curvatures always results in a rotationally symmetric shape, with regions of negative curvature close to the rim due to a geometric incompatibility. The competition with negative curvatures can lead to symmetric shapes when extrinsic curvature dominates, or to both positive and negative curvatures distributed across the surface. By considering the residual strains in these shapes, we found that the average absolute strain in the system scales linearly with the total curvature of the pattern, both before and after relaxation. Meanwhile, the average residual strain in an individual layer normalized by the total spontaneous strain in the system scales with $\alpha^2 K_\text{total}$. The three states emerging from the competition with saddle or anti-cone correspond to cleanly separated regions on the collapsed curves obtained by rescaling with $\alpha^2 K_\text{total}$. Finally, we have shown a number of interesting examples of systems with competition between two active, shape-programmed layers, each with distinct patterns, where even richer behavior is possible. 

Since there is yet no straightforward way to predict the shape formed by competing patterns and the significant residual stresses associated to them, we focused on simple pairs of patterns in order to have a better understanding of the equilibrated shapes and a better hope that these simple cases could provide the beginnings of a kind of 'vocabulary' and 'grammar' for layered shape-programming. 
We consistently observed that extrinsic curvature dominates for thick sheets and when the strain gradient is low, regardless of positive or negative curvature in the competing layers. 
Further, competitions between active positive curvature shapes result in symmetric shapes in space, although non-trivial negative Gaussian curvature can surprisingly still arise here as well. 
Similarly, shapes generated by active negative curvature layers and passive layers also can exhibit domains with both positive curvature and negative curvature.
Finally, using the intrinsic and extrinsic curvature properties of spontaneous strain-mismatched bilayers, we demonstrated the ability to program shapes inspired by biological systems, with the ventral furrow and multicusped tooth as examples. 

Investigating the residual strains left over in these systems once they are at their equilibrated configuration, we found that competitions of this type leave a large fraction of spontaneous strain still unresolved as compared to the total initial spontaneous strain. 
These residual spontaneous strains in the system could provide the stress profiles required to drive designed multi-stable branches in engineered shape-shifting materials. 
In biology, these patterns of residual spontaneous strains and their attendant stresses could be leveraged as a kind of signaling mechanism to initiate subsequent stages of morphogenesis, to control differential growth, or to guide functional adaptation. Indeed, in these contexts, internal stress read-outs through mechanosensitive signaling pathways are already known to occur. In future work, it will be interesting to consider feedback couplings between and among diffusing morphogen gradients, the developing shape of the bilayer, and the mechanosensitive signaling channels that react to the residual stresses. Indeed, it is possible that the mechanics of active bilayer shaping together with these feedbacks could lead to entirely self-organized shape motifs.

In this work we focused on the case of an active shape-programmed bilayer disk with free boundary conditions. 
However, morphogenesis in real biological systems typically occurs under non-trivial boundary constraints, while in engineering and device design active, shape-programmed components will almost always be at least partially contiguous with passive material.
Tissues and organs also often exhibit different topologies, such as spherical or tubular, and in the ventral furrow example, the flat regions away from the furrow cannot be maintained without imposing boundary constraints. 
Understanding incompatible spontaneous strains in the presence of such boundary conditions remains a rich and largely unexplored direction for future research. 
Ultimately, layered shape-programming holds significant promise for understanding complex tissue morphogenesis in biology and for device design applications in engineering. We are only just beginning to peel back the layers of potential held by these material systems.

\begin{acknowledgments}
The authors thank Alf Honigmann for fruitful discussions and insightful suggestions regarding the project, as well as Stephan W. Grill and Lior Moneta for a critical reading of the manuscript and valuable feedback. 
C.D.M. and W.Y.Y. acknowledge financial support by the VolkswagenStiftung within the Life? initiative.
\end{acknowledgments}

%\bibliographystyle{apsrev4-2}
%\bibliography{citation1,citation3,citation2,citation5}
\bibliography{citation_all}
\clearpage
\setcounter{section}{0}
\renewcommand{\thesection}{S.\Roman{section}}
\begin{center}
\textbf{\large Supplemental Materials}
\end{center}

\setcounter{equation}{0}
\setcounter{figure}{0}
\setcounter{table}{0}
\setcounter{page}{1}

\makeatletter
\renewcommand{\theequation}{S\arabic{equation}}
\renewcommand{\thefigure}{S\arabic{figure}}

\renewcommand{\theHequation}{Supplement.\theequation}
\renewcommand{\theHfigure}{Supplement.\thefigure}

\section{Simulation method}\label{S: simulation}
The simulations carried out here are based on a spring lattice model. Points are sampled on two layers of disks using Poisson disk sampling and further relaxed via Lloyd's relaxation to eliminate lattice symmetries and hence avoid a predetermined symmetry-breaking axis. The in-plane points are then triangulated using Delaunay triangulation, whose edges then become the in-plane springs. The inter-layer springs are constructed by connecting each point on one surface to its three nearest neighbors on the opposite surface. The spontaneous strain is programmed by defining the target length of each spring as the swelled length, calculated by integrating the swelling function along the path connecting the two endpoints. A linear gradient of swelling is assumed across the thickness. The entire system is then relaxed via force balance using \texttt{OpenFPM}~\cite{INCARDONA2019,Singh2025}, resulting in the final equilibrium configuration. For each simulated configuration, the pointwise curvature profile is computed using \texttt{scale-dependent quadric fitting} via \texttt{pymeshlab}~\cite{pymeshlab}. 

\section{Swelling functions for the simple shapes}\label{S:swelling_functions}
\subsection{Spherical cap and saddle}
Consider a spherical cap (Fig.~\ref{fig: spherica_cap_schematic}) of curvature radius $\gamma_K R_0$ and angular extent $\phi_0$. The in-material radius and the radius of circles at geodesic distance $r$ are $\gamma_K R_0\phi_0(r)$ and $\gamma_K R_0\sin\phi_0(r)$, respectively. Thus, the in-material radius gives the equation
\begin{align*}
\int_0^r \eta(r')\,dr' &= \gamma_K R_0\,\phi_0(r)
\\
\eta(r) &= \gamma_K R_0\,\frac{d\phi_0}{dr},
\end{align*}
and geodesic distance $r$ gives
\begin{align*}
r\,\eta(r) &= \gamma_K R_0 \sin\phi_0(r).
\end{align*}
These combine into
\[
\frac{d\phi_0}{dr} = \frac{\sin\phi_0}{r},
\]
with solution
\[
\phi_0(r) = 2\tan^{-1}(r/a),\qquad a\in\mathbb{R}.
\]
Thus the swelling function is
\begin{equation}
\eta_{s+}(r) = \frac{2a\gamma_K R_0}{a^2+r^2}, \qquad
\phi_0 \leq \pi.
\end{equation}
The expression for saddle shape with an equal magnitude of curvature can be derived in a similar way, but with hyperbolic geometry. 
The differential equation is
\begin{align*}
\frac{d\phi_0}{dr} = \frac{\sinh\phi_0}{r}
\end{align*}
and solving results in the swelling function
\begin{equation}
\eta_{s-}(r) = \frac{2a\gamma_K R_0}{a^2-r^2},\qquad a>0.
\end{equation}

\subsection{Cone and anti-cone}
The cone swelling function follows from the constant ratio between circumference and in-material radius, equal to $\sin\phi$ for a cone of opening angle $\phi$:
\[
\frac{r\,\eta(r)}{\rho(r)} = \sin\phi.
\]
Writing $\rho'(r)=\eta(r)$ gives the differential equation
\[
r\,\eta'(r) + \eta(r) = \sin\phi\,\eta(r),
\]
which has the solution
\begin{equation}
\eta_{c+}(r) = \left(\frac{r}{R'}\right)^{\sin\phi - 1}.
\end{equation}
The angle deficit of a cone is $2\pi (1-\sin\phi)$. 
An anti-cone with the same magnitude of Gaussian curvature requires an angle surplus of the same value. 
Hence, the circumference to in-material radius ratio requires a total angle of $2\pi(2-\sin\phi)$, and solving the corresponding differential equation results in the swelling function
\begin{equation}
\eta_{a-}(r) = \left(\frac{r}{R'}\right)^{1-\sin\phi}.
\end{equation}

\section{Derivation of the energetics}\label{S.1}
Consider an isotropically swelling bilayer, with the top and bottom surface uniformly swelling with a factor of $b$ and $a$ ($b>a$) respectively. The swelling function of the thick sheet can be described by
\begin{equation}\label{eq: uniform_swell_function_3D}
\eta(z) = \eta_-\frac{z}{h} + \eta_+, \quad \eta_- = b-a,\, \eta_+ = \frac{a+b}{2},
\end{equation}
on a disk with radius $R_0$ and thickness from $-h/2$ to $h/2$.

\subsection{Spherical cap}
\begin{figure*} [htbp]
\begin{center}
\includegraphics[width = 0.3\textwidth]{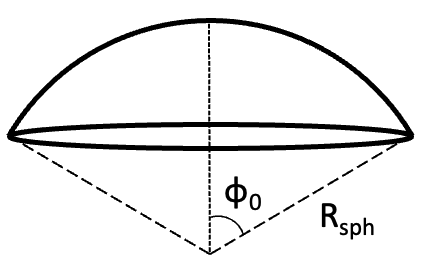}
\caption{Schematic of the bilayer in a spherical cap shape.}
\label{fig: spherica_cap_schematic}
\end{center}
\end{figure*}

Suppose the mismatch leads to a deformation into a spherical cap, as shown in Fig.~\ref{fig: spherica_cap_schematic}, this spherical cap is uniquely determined by the parameters $\phi_0$ and $R_{s+}$.
The bending energy is constant everywhere on the surface, and the energy density if $e_b = \frac{h^3}{24}\frac{E}{1-\nu}\frac{1}{R_{s+}^2}$.
The total bending energy is the product of the energy density and the area of the surface

\begin{equation}
\begin{split}
E_b &= e_b \cdot \int_0^{\phi_0} 2\pi R_{s+}^2 \sin\phi \,d\phi \\
    &= \frac{\pi h^3 E (1-\cos\phi_0)}{12(1-\nu^2)}
\end{split}
\end{equation}

The stretching energy can be evaluated by decomposing the compression density in the three directions
\begin{widetext}
\begin{align*}
\Lambda_{rr}(z) = \frac{(R_\text{sph}+z)\phi_0}{R_0\eta(z)}\, ,\quad
\Lambda_{\phi\phi}(r,z) = \frac{(R_\text{sph}+z)\sin(r\phi_0/R_0)}{r\eta(z)}
\, ,\quad
\Lambda_{hh} = 1.
\end{align*} 
\end{widetext}
By integrating the height dependent stretch energy density over the volume, 
\begin{widetext}
\begin{align*}
&\text{Stretch energy} \\
&= \frac{1}{2}E\int_0^{R_0} \int_{-h/2}^{h/2} \left[\left(\Lambda_{rr}(z)-1\right)^2 + \left(\Lambda_{\phi\phi}(r,z)\right)^2 + \left(\Lambda_{hh}-1\right)^2 \right] 2\pi r\eta(z) \cdot \eta(z)\,dz\,dr 
\end{align*}
\end{widetext}
The three terms can be evaluates separately:
\begin{widetext}
\begin{align*}
\text{First term} = &\frac{1}{2}E\int_0^{R_0} \int_{-h/2}^{h/2} \left(\Lambda_{rr}(z)-1\right)^2 2\pi r\eta(z) \cdot \eta(z)\, dr\, dz 
\\
=& \frac{1}{2}E\cdot 2\pi \int_0^{R_0}\int_{-h/2}^{h/2} r\left[ \left(\frac{R_\text{sph}\phi_0}{R_0} - \eta_+\right)+z\left(\frac{\phi_0}{R_0} - \frac{\eta_-}{h}\right)\right]^2\,dz\,dr
\\
=& \frac{\pi E R_0^2}{2}\left[h\left(\frac{R_\text{sph}\phi_0}{R_0} - \eta_+ \right)^2 + \frac{h^3}{12}\left(\frac{\phi_0}{R_0} - \frac{\eta_-}{h}\right)^2\right]
\\
\text{Second term} =& \frac{1}{2}E\int_0^{R_0} \int_{-h/2}^{h/2} \left(\Lambda_{\phi\phi}(z)-1\right)^2 2\pi r\eta(z) \cdot \eta(z)\, dr\, dz 
\\=& \frac{1}{2}E\cdot 2\pi \int_0^{R_0}\int_{-h/2}^{h/2} r
\left[ 
\left(\frac{R_\text{Rsph}\sin(r\phi_0/R_0)}{r} - \eta_+\right)
+
z\left(\frac{\sin(r\phi_0/R_0)}{r}-\frac{\eta_-}{h}\right)
\right]^2\,dz\,dr
\\=&
\pi E\Big\{
\frac{R_0^2}{2}\left(h\eta_+^2+\frac{h\eta_-^2}{12}\right) + 
\left(\frac{R_0\left[\cos(\phi_0)-1\right]}{\phi_0}\right) \left(2h\eta_+ R_\text{sph} + \frac{\eta_- h^2}{6}\right) + \\
&\qquad\qquad\qquad\qquad\qquad\qquad\qquad\qquad\qquad\qquad\left( hR_\text{sph}^2 + \frac{h^3}{12}\right)\int_0^{R_0} \frac{\sin^2\left(r \phi_0/R_0\right)}{r}\,dr
\Big\}
\end{align*}
\end{widetext}

The integral in the form $\int \frac{\sin^2(ar)}{r} dr $ can be evaluated:
\begin{align*}
&\int \frac{\sin^2(ar)}{r} dr = \frac{\text{Ci}(2 a r)}{2}+\frac{\log (r)}{2},\\
&\int_0^{R_0}  \frac{\sin^2(ar)}{r} dr =\frac{1}{2} [-\text{Ci}(2 a R_0)+\log (2 a R_0)+\gamma ]
\end{align*}
To extract the dependence on the aspect ratio $\alpha = 2R_0/h$ of the sheet with volume $V = \pi R_0^2 h$, the energy is rewritten by factorizing out the volume, and substitute $\alpha = 2R_0/h$ and $\gamma_{s+} = R_{s+}/R_0$. 
The total energy is
\begin{widetext}
\begin{equation}\label{eq: sph_energy}
 \begin{aligned}
\text{Total energy} =& \frac{EV}{1-\nu^2}\frac{(1-\cos\phi_0)}{3\alpha^2} +\\
&\frac{VE}{2}\Big[
2\left(\eta_+^2 + \frac{\eta_-^2}{12}\right)+
\left(2\eta_+\gamma_{s+} + \frac{\eta_-}{3\alpha} \right) \left(\frac{2(\cos\phi_0-1)}{\phi_0}-\phi_0\right) +\\
&\qquad\qquad\qquad\qquad\left(\gamma_{s+}^2 + \frac{1}{3\alpha^2}\right)\left(\phi_0^2 -\text{Ci}(2 \phi_0)+\log (2 \phi_0)+\gamma 
\right)
\Big]
 \end{aligned}
\end{equation}
\end{widetext}
%%%%%%%%%%%%%%%%%
\subsection{Cylinder}
\begin{figure*}[htbp]
\begin{center}
\includegraphics[width=\textwidth]{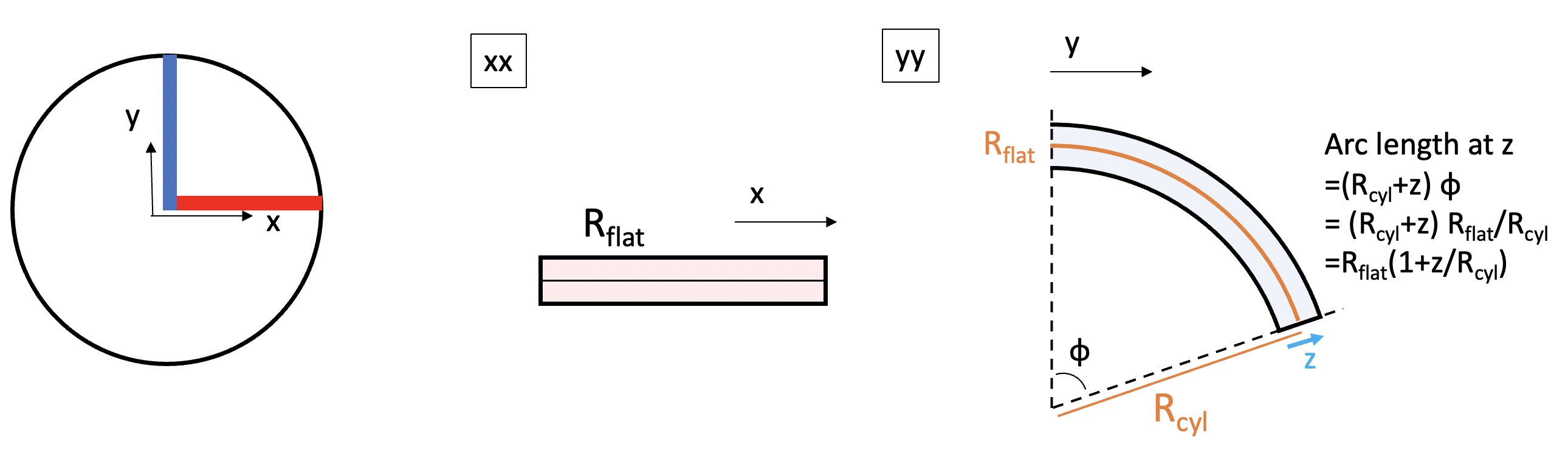}
\caption{Schematic of a cylindrical surface curling in the y axis. Left: the x and y axis defined on a disk, center: cross section of the curled disk along x, right: cross section along y }
\label{default}
\end{center}
\end{figure*}

The sheet undergoes uniform expansion or shrinkage from the initial radius $R_0$ to a new radius $R_\text{flat}$ and next rolls in one direction with radius of curvature $1/R_\text{cyl}$. 
The configuration has a mean curvature $\frac{1}{2} \cdot (0 + \frac{1}{R_\text{cyl}}) = \frac{1}{2R_\text{cyl}}$.
Hence, bending energy density and bending energy are:
\begin{align*}
e_b &= \frac{1}{2}\frac{E}{1-\nu_\text{el}^2}\frac{h^3}{12}\frac{1}{4R_\text{cyl}^2} \\
&= \frac{E h^3}{96R_\text{cyl}^2(1-\nu^2)}
\\
\text{Bending energy} &= \pi R_0^2 \cdot e_b \\
&=\frac{\pi E h^3R_0^2}{96R_\text{cyl}^2(1-\nu^2)}
\end{align*}
Since symmetry is broken in one direction, we write the stretch and bend energy in Cartesian coordinates.
Assume the sheet curls along $y$, 
\begin{align*}
&\Lambda_{xx}(x,y,z) = \frac{R_\text{flat}}{\eta(z)R_0} = \frac{\eta(0)R_0}{\eta(z)R_0} = \frac{\eta(0)}{\eta(z)}
\\
&\Lambda_{yy}(x,y,z) = \frac{R_\text{flat}(1+z/R_\text{cyl})}{\eta(z)R_0} = \frac{\eta(0)(1+z/R_\text{cyl})}{\eta(z)}
\\
&\Lambda_{zz} = 1
\end{align*}

Integrating stretch energy over a quarter of the disk
\begin{widetext}
\begin{align*}
&\frac{1}{4}\cdot \text{stretching energy} 
\\=& \frac{E}{2}\int^{R_0}_0\int^{\sqrt{R_0^2-y^2}}_0\int^{\frac{h}{2}}_{-\frac{h}{2}} \left(\frac{x^2}{x^2+y^2}(\Lambda_{xx}-1)^2+\frac{y^2}{x^2+y^2}(\Lambda_{yy}-1)^2 + (\Lambda_{zz}-1)^2\right)\eta(z)^2 \,dz\,dx\,dy
\\
=&\frac{E}{2}\int^{R_0}_0\int^{\sqrt{R_0^2-y^2}}_0\int^{\frac{h}{2}}_{-\frac{h}{2}} z^2 \left(
\frac{x^2}{x^2+y^2}\frac{\eta_-^2}{h^2} +
\frac{y^2}{x^2+y^2} \left(\frac{\eta_0}{R_\text{cyl}}-\frac{\eta_-}{h}\right)^2 
\right)\,dz\,dx\,dy
\\
=&\frac{E}{2}\frac{\pi R_0^2}{8}\left[\frac{\eta_-^2 h}{12}
+\frac{h^3}{12}\left(\frac{\eta_0}{R_\text{cyl}} - \frac{\eta_-}{h}\right)^2 
\right]
\end{align*}
\end{widetext}
Again, we rewrite the energy in terms of the volume $V$, aspect ratio $\alpha$, and the dimensionless curvature parameter $\gamma_c = R_\text{cyl}/R_0$.
The total energy becomes
\begin{widetext}
\begin{equation}\label{eq: cyl_energy}
\text{Total energy} = \frac{EV}{24(1-\nu^2)}\frac{1}{\gamma_c^2\alpha^2} + \frac{EV}{48}\left(\eta_-^2+\left(\frac{2\eta_0}{\gamma_c\alpha}-\eta_-\right)^2\right)^2
\end{equation}
\end{widetext}
For the simulations performed, the top surface has a uniform shrinkage of $1-d$ and the bottom surface is passive.
Therefore, $\eta_+ = 1-d/2$ and $\eta_- = d$ can be substituted into the energy expressions is Eq.~\ref{eq: sph_energy}  and ~\ref{eq: cyl_energy}, and they can be rewritten as follows:
\begin{widetext}
\begin{equation}\label{eq: sph_energy_d}
 \begin{aligned}
E_s = \frac{EV}{1-\nu^2}\frac{(1-\cos\phi_0)}{3\alpha^2} +
\frac{VE}{2}\Big[
2\left(1-d+\frac{d^2}{3}\right)+
\left((2-d)\gamma_{s+} + \frac{d}{3\alpha} \right) \left(\frac{2(\cos\phi_0-1)}{\phi_0}-\phi_0\right) +\\
\left(\gamma_{s+}^2 + \frac{1}{3\alpha^2}\right)\left(\phi_0^2 -\text{Ci}(2 \phi_0)+\log (2 \phi_0)+\gamma 
\right)
\Big]
\end{aligned}
\end{equation}
\end{widetext}
\begin{equation}\label{eq: cyl_energy_d}
\begin{split}
E_c =& \frac{EV}{24(1-\nu^2)}\frac{1}{\gamma_c^2\alpha^2} + \\
&\frac{EV}{48}\left(d^2+\left(\frac{2-d}{\gamma_c\alpha}-d\right)^2\right)^2
\end{split}
\end{equation}

\section{Dependence of initial and final residual strains on aspect ratio}\label{S.strain_R0}
\begin{figure*} [htbp]
\begin{center}
\includegraphics[width=\textwidth]{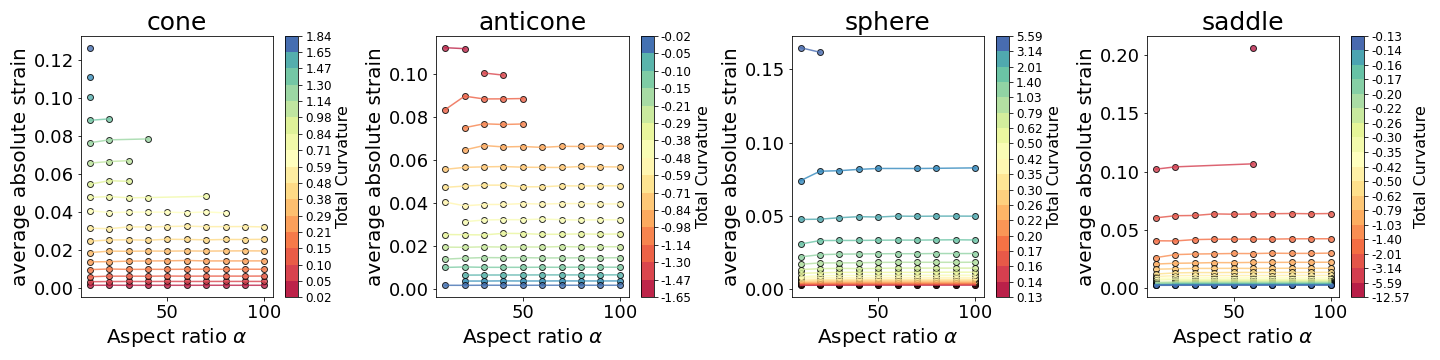}
\caption{The initial average absolute strain over all the springs plotted against the aspect ratio of the starting sheet.}
\label{strain_initial}
\end{center}
\end{figure*}

\begin{figure*} [htbp]
\begin{center}
\includegraphics[width=\textwidth]{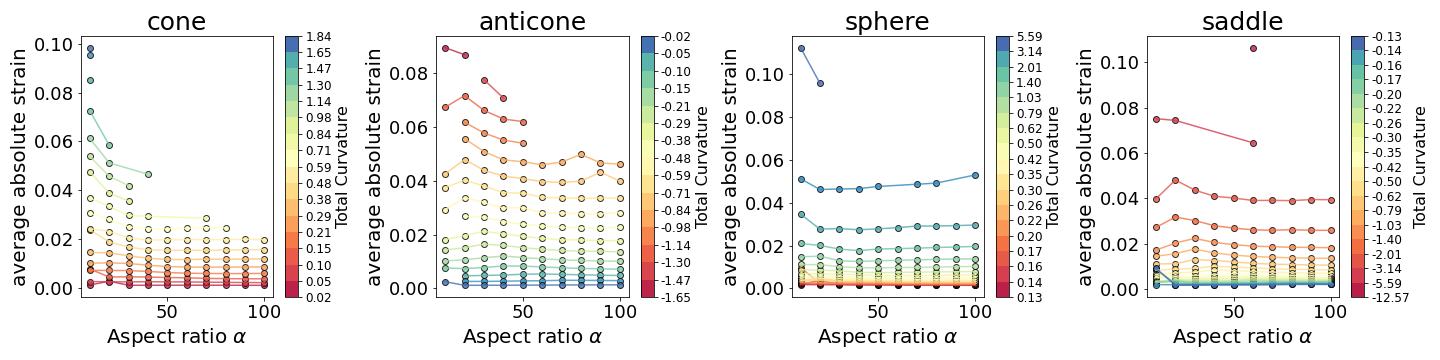}
\caption{The final average absolute strain over all the springs plotted against the aspect ratio of the starting sheet.}
\label{strain_final}
\end{center}
\end{figure*}
The initial average absolute strain has almost no dependence on the aspect ratio, mean value of the strains across different aspect ratio for total curvatures is taken to plot the figure Fig.~\ref{Fig: strain}. 
The initial average absolute strain shows no dependence on the aspect ratio (Fig.~\ref{strain_initial}), and the final strains are roughly constant, except at lower values of aspect ratio(Fig.~\ref{strain_final}). 

\section{Final residual strains on individual surface}\label{S: normalized_strain}
\begin{figure*} [htbp]
\begin{center}
\includegraphics[width=\textwidth]{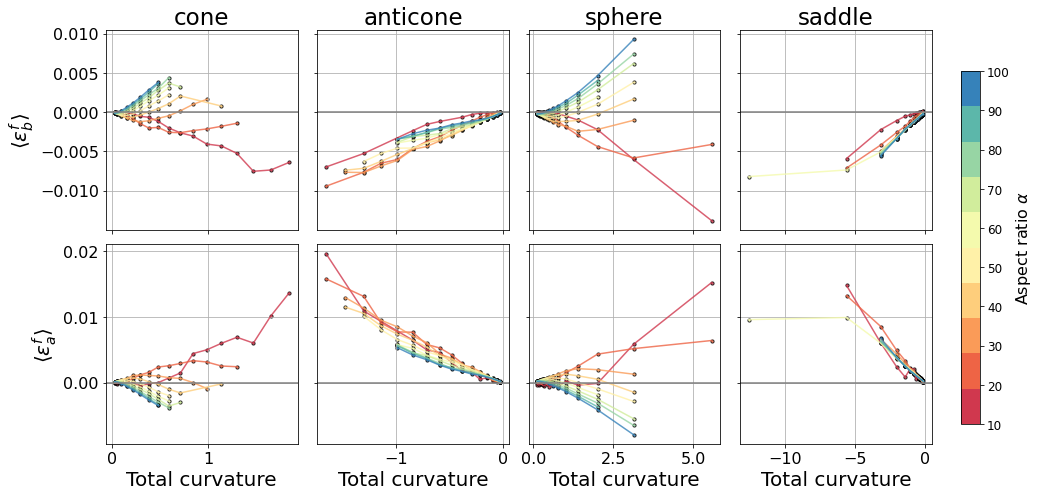}
\caption{The final average strain plotted against total target curvature of the active surface, colored by the aspect ratios. }
\label{strain_initial}
\end{center}
\end{figure*}

This plot shows the residual strains of the apical (active) layer and the basal (passive) layer plotted against the total curvature programmed on the active surface. The final average strain on the two layers appear to be opposite, and have a monotonic trends in the hyperbolic shapes (anti-cone and saddle). The residual strains contributions on the two layers for parabolic shapes (cone and sphere) is affected by both the total target curvature and the aspect ratio.

\section{Curve collapses}\label{S: curve_collapse}
\begin{figure*} [htbp]
\begin{center}
\includegraphics[width=0.9\textwidth]{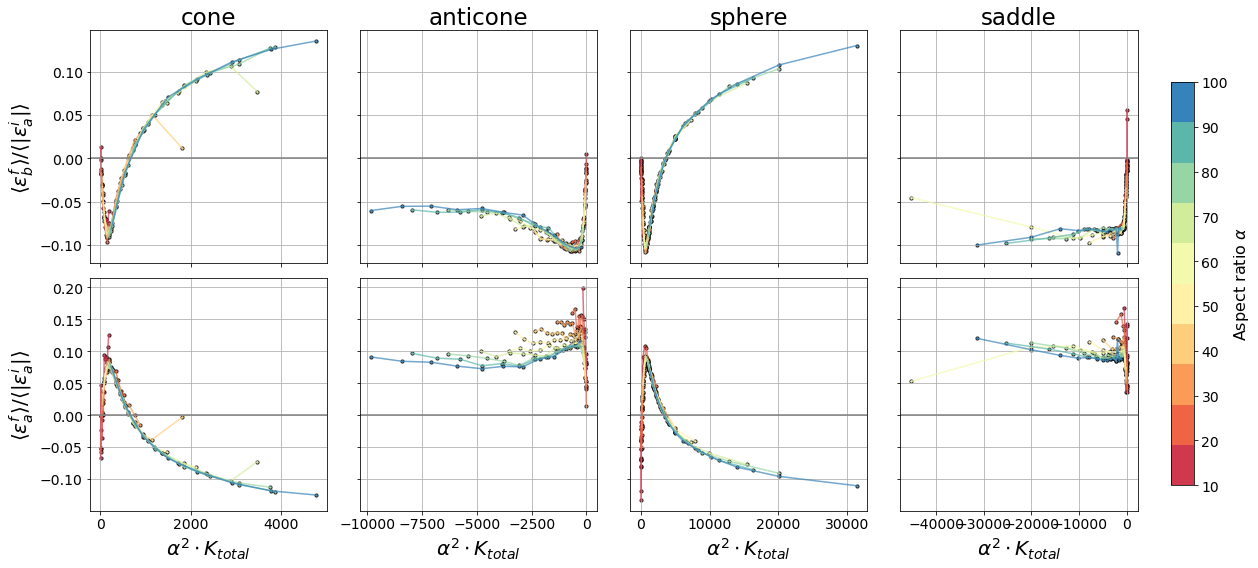}
\includegraphics[width=0.9\textwidth]{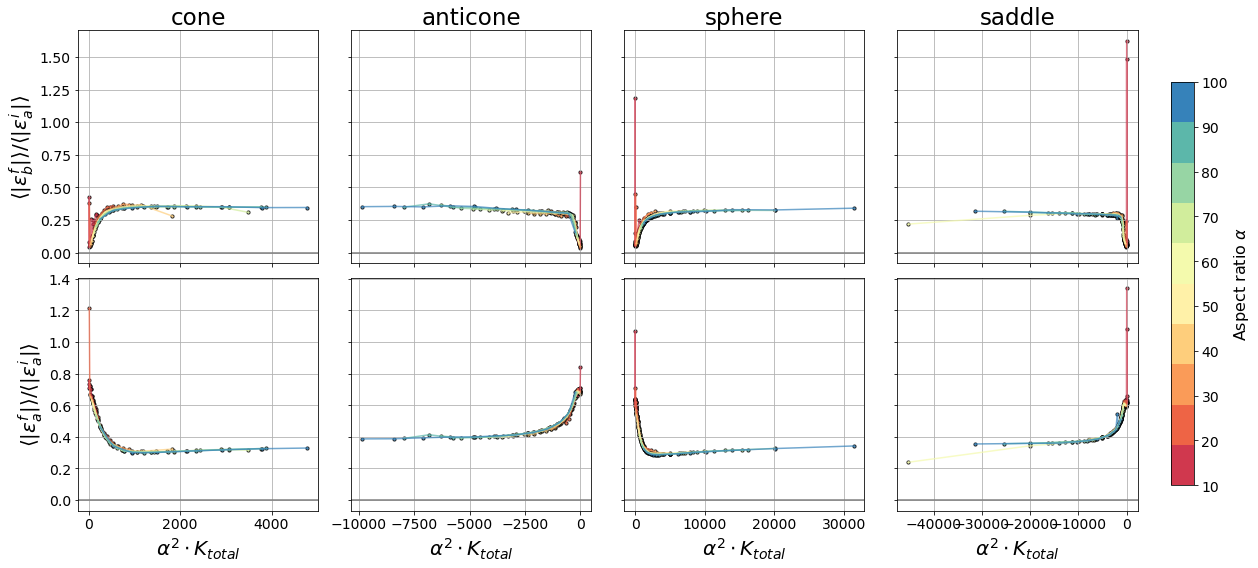}
\caption{Top: the average absolute residual strain of each surface normalized by the average absolute initial strain plotted against $\alpha^2K_\text{total}$. Bottom: The average absolute residual strain of each surface normalized by the average absolute initial strain plotted against $\alpha^2K_\text{total}$.}
\label{abs_collapse}
\end{center}
\end{figure*}

The strain quantities experiencing curve collapses when plotted against $\alpha^2 K_\text{total}$. 
In addition to the average final strain normalized by the initial absolute strain, the normalized average absolute final strain also shows this behavior. 
These two plots for all four competitions are shown in Fig.~\ref{abs_collapse}. 
For both the signed and the absolute valued plots, the programmed shapes with the same sign appear to have a similar trend. 

\begin{figure*} [htbp]
\begin{center}
\includegraphics[width=0.7\textwidth]{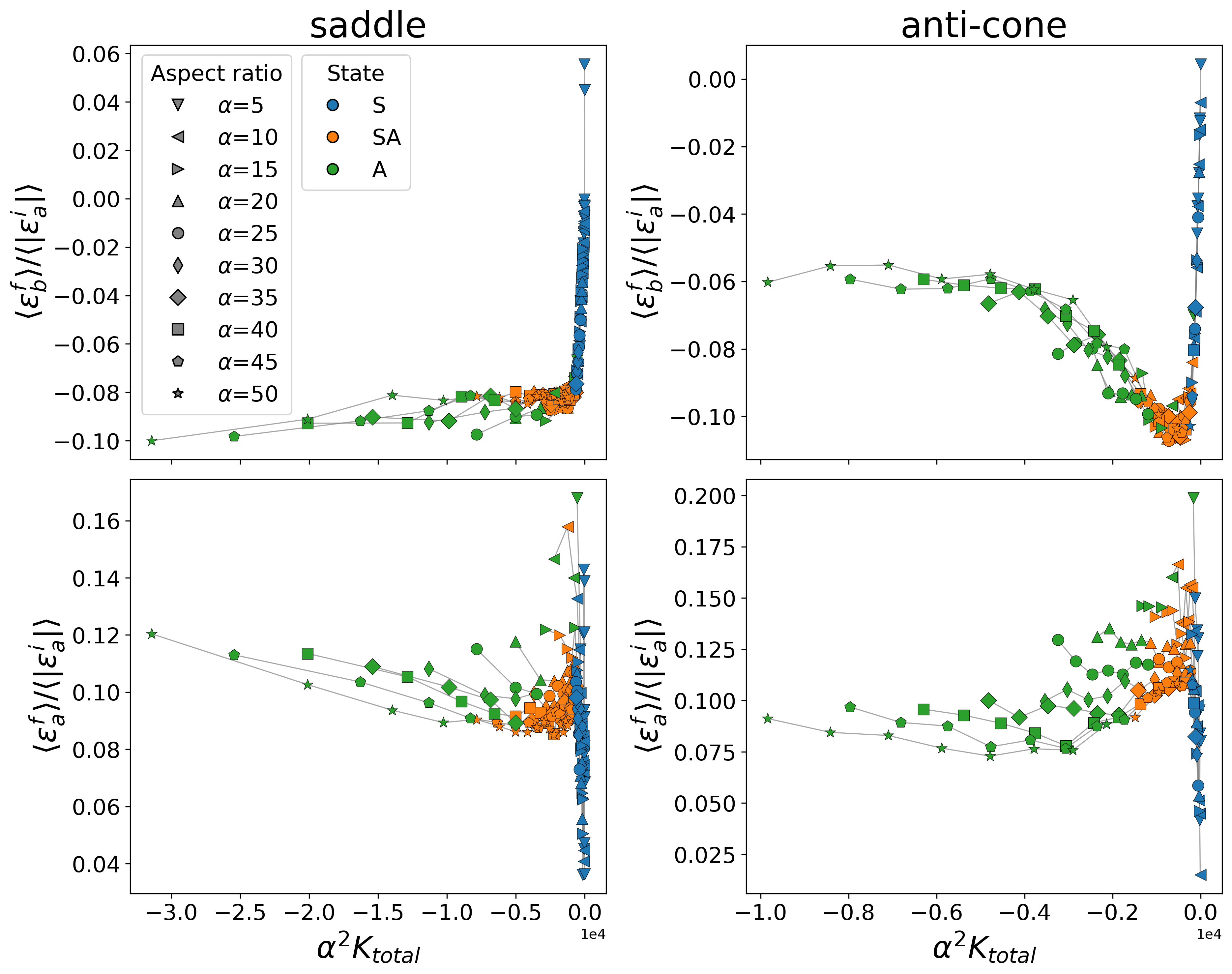}
\caption{The full version of Fig.~\ref{Fig: strain}(d) with curves of all aspect ratios plotted.}
\label{curve_collapse_all}
\end{center}
\end{figure*}

\section{Spatial profiles of residual strains}\label{S: distribution}
\begin{figure*} [htbp]
\begin{center}
\includegraphics[width=\textwidth]{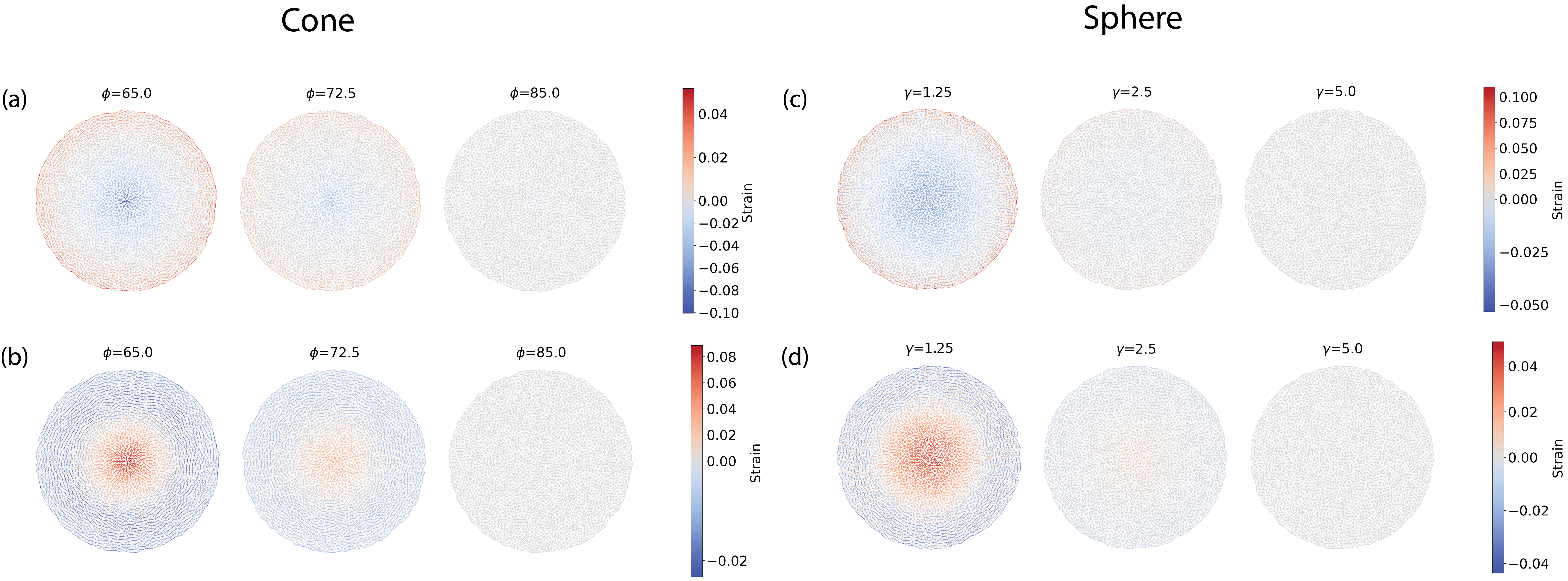}
\caption{The residual strain profiles of the positive curvature shapes competition for three curvature values each. The visualization of Gaussian curvature are normalized across the three curvature parameters. }
\label{strain_profile_positive}
\end{center}
\end{figure*}
The Residual strain profiles of the competitions with cone and sphere shapes. 
In contrast to the profiles fo saddle and anti-cone (Fig.~\ref{Fig: strain}a), the distribution for these two shapes are symmetric.

\section{Competitions between shapes}\label{S: competition}
This section presents the resulting shapes of competing patterns with nontrivial 
Gaussian curvature. To isolate the effects of competition between curvature 
distributions, the target total areas of the two layers are chosen to be equal. 
Because the curvature distributions differ for the cone, anti-cone, spherical cap, 
and saddle, we compare these shapes by controlling the total curvature on the 
surface. 

In the following plots, the total curvature is represented by the angle parameter 
for the cone and anti-cone. A curvature parameter of $\beta^\circ$ ($\beta = 60, 65, 70, 80$) induces a total 
absolute curvature of $2\pi (1-\sin\phi)$; the corresponding dimensionless radius 
of curvature, $\gamma$, can then be determined since the total curvature of the 
spherical cap or saddle surface is defined as $\pi/\gamma^2$.
\begin{figure*} [htbp]
\begin{center}
\includegraphics[width=\textwidth]{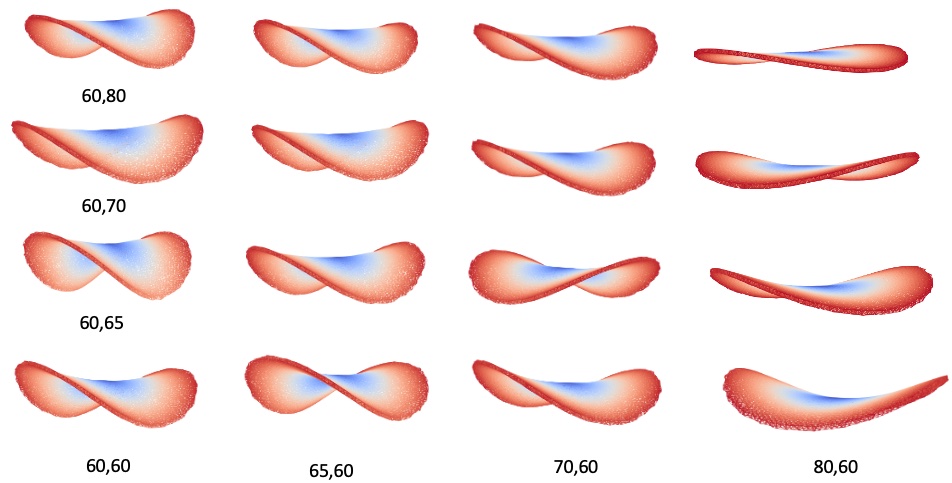}
\caption{The competition between anti-cone and saddle shapes. Resulting configurations are colored blue-to-red from the center of the initial reference disk to its edge for ease of visualization. The pair of parameters ($\beta_1, \beta_2$) are the curvature parameters of the anti-cone shape and the saddle shapes respectively. }
\label{}
\end{center}
\end{figure*}
\begin{figure*} [htbp]
\begin{center}
\includegraphics[width=\textwidth]{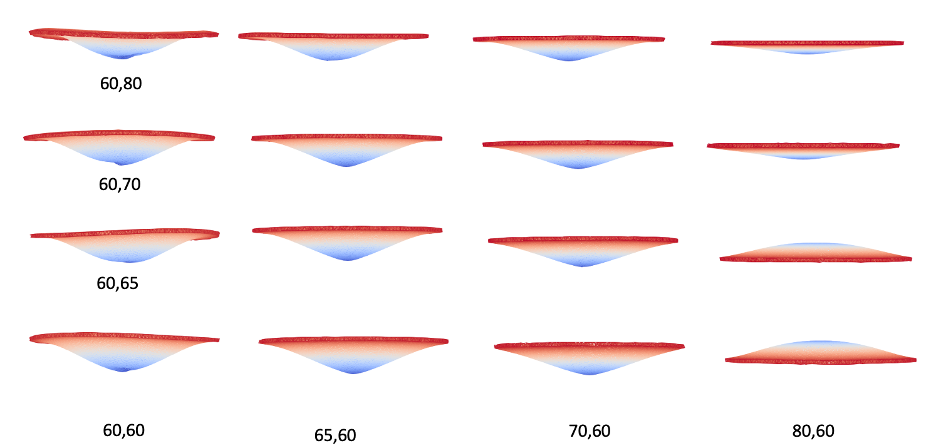}
\caption{The competition between cone and sphere shapes. Resulting configurations are colored blue-to-red from the center of the initial reference disk to its edge for ease of visualization. The pair of parameters ($\beta_1, \beta_2$) are the curvature parameters of the cone shape and the sphere shapes respectively. }
\label{}
\end{center}
\end{figure*}
\begin{figure*} [htbp]
\begin{center}
\includegraphics[width=\textwidth]{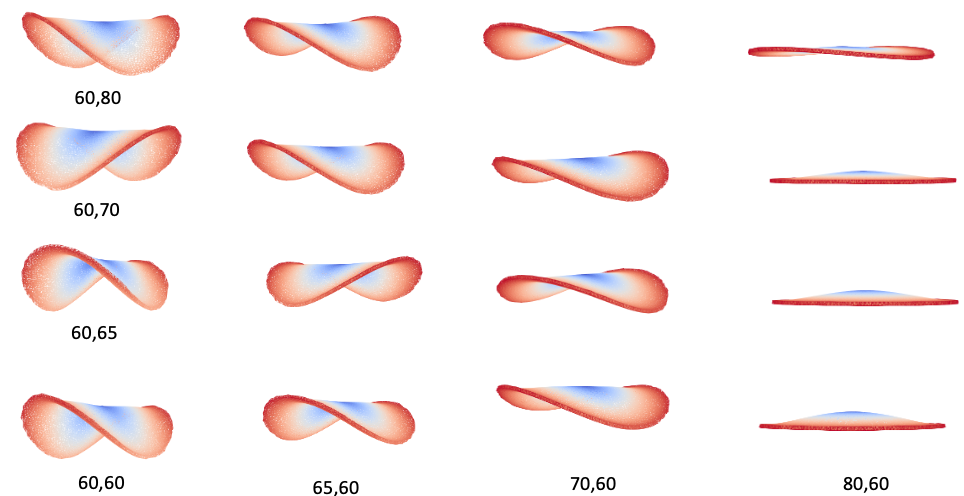}
\caption{The competition between anti-cone and sphere shapes. Resulting configurations are colored blue-to-red from the center of the initial reference disk to its edge for ease of visualization. The pair of parameters ($\beta_1, \beta_2$) are the curvature parameters of the anti-cone shape and the sphere shapes respectively. }
\label{}
\end{center}
\end{figure*}
\begin{figure*} [htbp]
\begin{center}
\includegraphics[width=\textwidth]{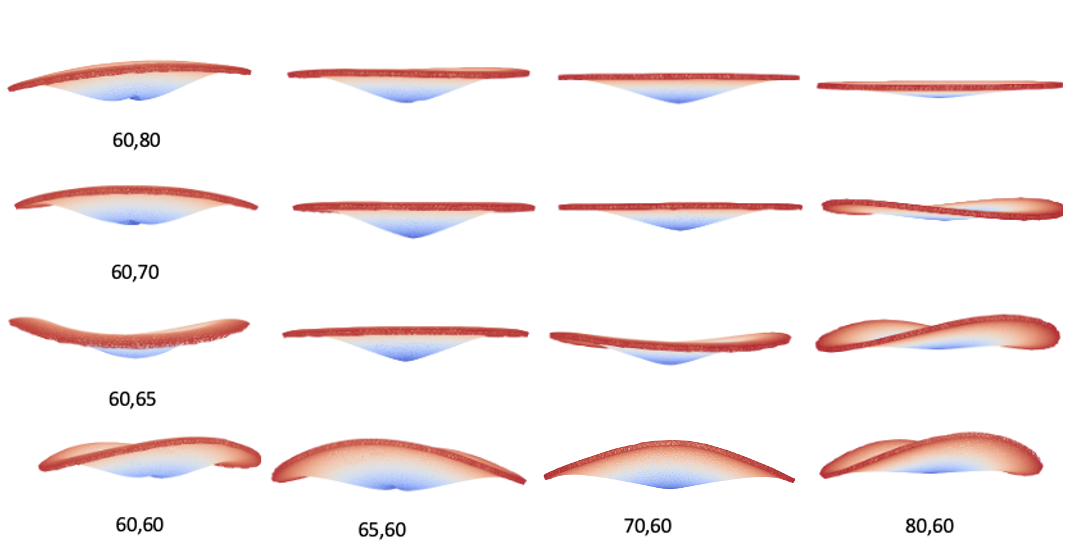}
\caption{The competition between cone and saddle shapes. Resulting configurations are colored blue-to-red from the center of the initial reference disk to its edge for ease of visualization. The pair of parameters ($\beta_1, \beta_2$) are the curvature parameters of the cone shape and the sphere shapes respectively. }
\label{}
\end{center}
\end{figure*}
\begin{figure*} [htbp]
\begin{center}
\includegraphics[width=\textwidth]{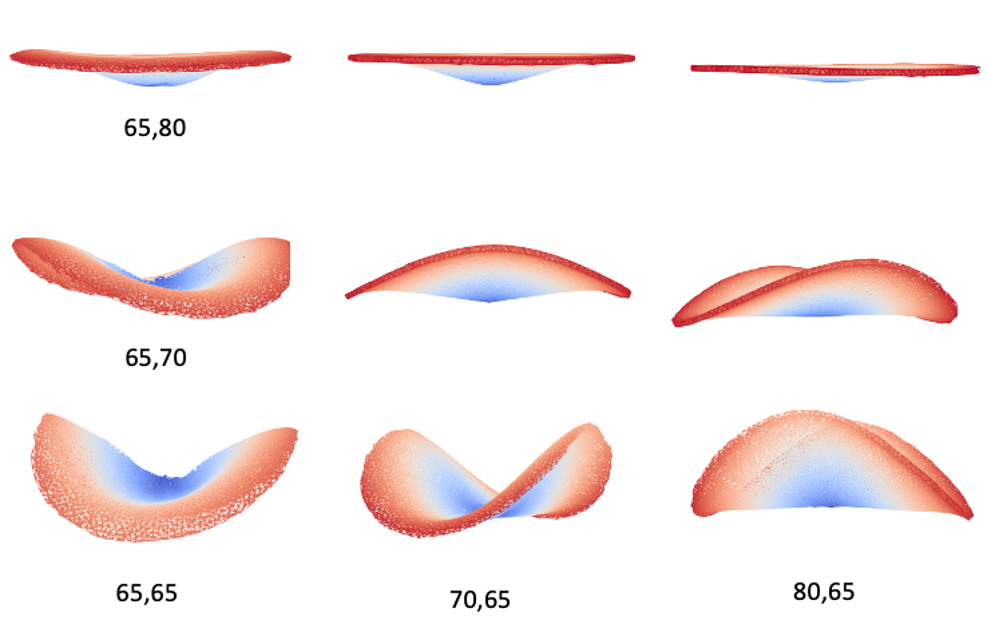}
\caption{The competition between cone and anti-cone shapes. Resulting configurations are colored blue-to-red from the center of the initial reference disk to its edge for ease of visualization. The pair of parameters ($\beta_1, \beta_2$) are the curvature parameters of the cone shape and the anti-cone shape respectively. }
\label{}
\end{center}
\end{figure*}
\begin{figure*} [htbp]
\begin{center}
\includegraphics[width=\textwidth]{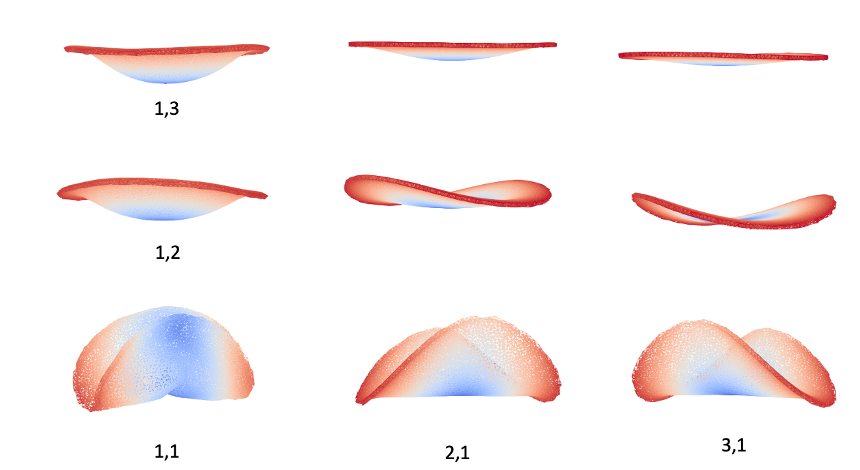}
\caption{The competition between sphere and saddle shapes. Resulting configurations are colored blue-to-red from the center of the initial reference disk to its edge for ease of visualization. The pair of parameters ($\beta_1, \beta_2$) are the curvature parameters of the sphere shape and the saddle shape respectively. }
\label{}
\end{center}
\end{figure*}

\end{document}